%% file: gauginomedv2.tex
\input harvmac
\input amssym
\input epsf
\input tables
\def\unit{\relax{\rm 1\kern-.26em I}}
\def\nada{\relax{\rm 0\kern-.30em l}}
\def\tilde{\widetilde}

%\draftmode

%\def\Omega{\rho,\sigma,\nu  }

\def\det{{\rm det}}

%% MACROS
\noblackbox
\def\IL{\relax{\rm I\kern-.18em L}}
\def\IH{\relax{\rm I\kern-.18em H}}
\def\IR{\relax{\rm I\kern-.18em R}}
\def\IC{\relax\hbox{$\inbar\kern-.3em{\rm C}$}}
\def\IZ{\relax\ifmmode\mathchoice
{\hbox{\cmss Z\kern-.4em Z}}{\hbox{\cmss Z\kern-.4em Z}}
{\lower.9pt\hbox{\cmsss Z\kern-.4em Z}} {\lower1.2pt\hbox{\cmsss
Z\kern-.4em Z}}\else{\cmss Z\kern-.4em Z}\fi}
\def\CM {{\cal M}}

\def\CO {{\cal O}}

\def\CA{{\cal A}}

%% MORE MACROS
\def\CM {{\cal M}}

\def\CO {{\cal O}}

\def\det{{\rm det}}
\def\Tr{{\rm Tr}}

\font\manual=manfnt \def\dbend{\lower3.5pt\hbox{\manual\char127}}

\def\IZ{\relax\ifmmode\mathchoice
{\hbox{\cmss Z\kern-.4em Z}}{\hbox{\cmss Z\kern-.4em Z}}
{\lower.9pt\hbox{\cmsss Z\kern-.4em Z}} {\lower1.2pt\hbox{\cmsss
Z\kern-.4em Z}}\else{\cmss Z\kern-.4em Z}\fi}

\def\rt2{\sqrt{2}}
\def\irt2{{1\over\sqrt{2}}}

%  \slashchar puts a slash through a character to represent contraction
%  with Dirac matrices. Use \not instead for negation of relations, and use
%  \hbar for hbar.
\def\slashchar#1{\setbox0=\hbox{$#1$}           % set a box for #1
   \dimen0=\wd0                                 % and get its size
   \setbox1=\hbox{/} \dimen1=\wd1               % get size of /
   \ifdim\dimen0>\dimen1                        % #1 is bigger
      \rlap{\hbox to \dimen0{\hfil/\hfil}}      % so center / in box
      #1                                        % and print #1
   \else                                        % / is bigger
      \rlap{\hbox to \dimen1{\hfil$#1$\hfil}}   % so center #1
      /                                         % and print /
   \fi}

\def\foursqr#1#2{{\vcenter{\vbox{
    \hrule height.#2pt
    \hbox{\vrule width.#2pt height#1pt \kern#1pt
    \vrule width.#2pt}
    \hrule height.#2pt
    \hrule height.#2pt
    \hbox{\vrule width.#2pt height#1pt \kern#1pt
    \vrule width.#2pt}
    \hrule height.#2pt
        \hrule height.#2pt
    \hbox{\vrule width.#2pt height#1pt \kern#1pt
    \vrule width.#2pt}
    \hrule height.#2pt
        \hrule height.#2pt
    \hbox{\vrule width.#2pt height#1pt \kern#1pt
    \vrule width.#2pt}
    \hrule height.#2pt}}}}
\def\psqr#1#2{{\vcenter{\vbox{\hrule height.#2pt
    \hbox{\vrule width.#2pt height#1pt \kern#1pt
    \vrule width.#2pt}
    \hrule height.#2pt \hrule height.#2pt
    \hbox{\vrule width.#2pt height#1pt \kern#1pt
    \vrule width.#2pt}
    \hrule height.#2pt}}}}
\def\sqr#1#2{{\vcenter{\vbox{\hrule height.#2pt
    \hbox{\vrule width.#2pt height#1pt \kern#1pt
    \vrule width.#2pt}
    \hrule height.#2pt}}}}

\def\figin{\epsfcheck\figin}\def\figins{\epsfcheck\figins}
\def\epsfcheck{\ifx\epsfbox\UnDeFiNeD
\message{(NO epsf.tex, FIGURES WILL BE IGNORED)}
\gdef\figin##1{\vskip2in}\gdef\figins##1{\hskip.5in}% blank space instead
\else\message{(FIGURES WILL BE INCLUDED)}%
\gdef\figin##1{##1}\gdef\figins##1{##1}\fi}
\def\DefWarn#1{}
\def\figinsert{\goodbreak\midinsert}
\def\ifig#1#2#3{\DefWarn#1\xdef#1{fig.~\the\figno}
\writedef{#1\leftbracket fig.\noexpand~\the\figno}%
\figinsert\figin{\centerline{#3}}\medskip\centerline{\vbox{\baselineskip12pt
\advance\hsize by -1truein\noindent\footnotefont{\bf
Fig.~\the\figno:\ } \it#2}}
\bigskip\endinsert\global\advance\figno by1}

\lref\gmreview{
  G.~F.~Giudice and R.~Rattazzi,
  ``Theories with gauge-mediated supersymmetry breaking,''
  Phys.\ Rept.\  {\bf 322}, 419 (1999)
  [arXiv:hep-ph/9801271].
  %%CITATION = PRPLC,322,419;%%
}

\lref\BrignoleCM{
  A.~Brignole, J.~A.~Casas, J.~R.~Espinosa and I.~Navarro,
  ``Low-scale supersymmetry breaking: Effective description, electroweak
  breaking and phenomenology,''
  Nucl.\ Phys.\  B {\bf 666}, 105 (2003)
  [arXiv:hep-ph/0301121].
  %%CITATION = NUPHA,B666,105;%%
  }

%\AffleckXZ
\lref\AffleckXZ{
  I.~Affleck, M.~Dine and N.~Seiberg,
  ``Dynamical Supersymmetry Breaking In Four-Dimensions And Its
  Phenomenological Implications,''
  Nucl.\ Phys.\  B {\bf 256}, 557 (1985).
  %%CITATION = NUPHA,B256,557;%%
}

%\DineYW
\lref\DineYW{
  M.~Dine and A.~E.~Nelson,
  ``Dynamical supersymmetry breaking at low-energies,''
  Phys.\ Rev.\  D {\bf 48}, 1277 (1993)
  [arXiv:hep-ph/9303230].
  %%CITATION = PHRVA,D48,1277;%%
}

%\DineVC
\lref\DineVC{
  M.~Dine, A.~E.~Nelson and Y.~Shirman,
  ``Low-Energy Dynamical Supersymmetry Breaking Simplified,''
  Phys.\ Rev.\  D {\bf 51}, 1362 (1995)
  [arXiv:hep-ph/9408384].
  %%CITATION = PHRVA,D51,1362;%%
}

%\LutyFK
\lref\LutyFK{
  M.~A.~Luty,
  ``Naive dimensional analysis and supersymmetry,''
  Phys.\ Rev.\  D {\bf 57}, 1531 (1998)
  [arXiv:hep-ph/9706235].
  %%CITATION = PHRVA,D57,1531;%%
}

%\DineAG
\lref\DineAG{
  M.~Dine, A.~E.~Nelson, Y.~Nir and Y.~Shirman,
  ``New tools for low-energy dynamical supersymmetry breaking,''
  Phys.\ Rev.\  D {\bf 53}, 2658 (1996)
  [arXiv:hep-ph/9507378].
  %%CITATION = PHRVA,D53,2658;%%
}

%\WittenKV
\lref\WittenKV{
  E.~Witten,
  ``Mass Hierarchies In Supersymmetric Theories,''
  Phys.\ Lett.\  B {\bf 105}, 267 (1981).
  %%CITATION = PHLTA,B105,267;%%
}

%\BanksMG
\lref\BanksMG{
  T.~Banks and V.~Kaplunovsky,
  ``Nosonomy Of An Upside Down Hierarchy Model. 1,''
  Nucl.\ Phys.\  B {\bf 211}, 529 (1983).
  %%CITATION = NUPHA,B211,529;%%
}
%\KaplunovskyYX
\lref\KaplunovskyYX{
  V.~Kaplunovsky,
  ``Nosonomy Of An Upside Down Hierarchy Model. 2,''
  Nucl.\ Phys.\  B {\bf 233}, 336 (1984).
  %%CITATION = NUPHA,B233,336;%%
}

%\DimopoulosGM
\lref\DimopoulosGM{
  S.~Dimopoulos and S.~Raby,
  ``Geometric Hierarchy,''
  Nucl.\ Phys.\  B {\bf 219}, 479 (1983).
  %%CITATION = NUPHA,B219,479;%%
}

%\DermisekQJ
\lref\DermisekQJ{
  R.~Dermisek, H.~D.~Kim and I.~W.~Kim,
  ``Mediation of supersymmetry breaking in gauge messenger models,''
  JHEP {\bf 0610}, 001 (2006)
  [arXiv:hep-ph/0607169].
  %%CITATION = JHEPA,0610,001;%%
}

%\DineGU%
\lref\DineGU{
  M.~Dine and W.~Fischler,
  ``A Phenomenological Model Of Particle Physics Based On Supersymmetry,''
  Phys.\ Lett.\  B {\bf 110}, 227 (1982).
  %%CITATION = PHLTA,B110,227;%%
}

%\NappiHM
\lref\NappiHM{
  C.~R.~Nappi and B.~A.~Ovrut,
  ``Supersymmetric Extension Of The SU(3) X SU(2) X U(1) Model,''
  Phys.\ Lett.\  B {\bf 113}, 175 (1982).
  %%CITATION = PHLTA,B113,175;%%
}

%\DimopoulosAU
\lref\DimopoulosAU{
  S.~Dimopoulos and S.~Raby,
  ``Supercolor,''
  Nucl.\ Phys.\  B {\bf 192}, 353 (1981).
  %%CITATION = NUPHA,B192,353;%%
}

%\DineZB
\lref\DineZB{
  M.~Dine and W.~Fischler,
  ``A Supersymmetric Gut,''
  Nucl.\ Phys.\  B {\bf 204}, 346 (1982).
  %%CITATION = NUPHA,B204,346;%%
}

%\AlvarezGaumeWY
\lref\AlvarezGaumeWY{
  L.~Alvarez-Gaume, M.~Claudson and M.~B.~Wise,
  ``Low-Energy Supersymmetry,''
  Nucl.\ Phys.\  B {\bf 207}, 96 (1982).
  %%CITATION = NUPHA,B207,96;%%
}

\lref\tobenomura{
  Y.~Nomura and K.~Tobe,
  ``Phenomenological aspects of a direct-transmission model of dynamical
  supersymmetry breaking with the gravitino mass m(3/2) $<$ 1-keV,''
  Phys.\ Rev.\  D {\bf 58}, 055002 (1998)
  [arXiv:hep-ph/9708377].
}

\lref\IzawaGS{
  K.~I.~Izawa, Y.~Nomura, K.~Tobe and T.~Yanagida,
  ``Direct-transmission models of dynamical supersymmetry breaking,''
  Phys.\ Rev.\  D {\bf 56}, 2886 (1997)
  [arXiv:hep-ph/9705228].
  %%CITATION = PHRVA,D56,2886;%%
}

%\CheungES
\lref\CheungES{
  C.~Cheung, A.~L.~Fitzpatrick and D.~Shih,
  ``(Extra)Ordinary Gauge Mediation,''
  arXiv:0710.3585 [hep-ph].
  %%CITATION = ARXIV:0710.3585;%%
}

\lref\hiddenren{
  A.~G.~Cohen, T.~S.~Roy and M.~Schmaltz,
  ``Hidden sector renormalization of MSSM scalar masses,''
  JHEP {\bf 0702}, 027 (2007)
  [arXiv:hep-ph/0612100].
  %%CITATION = JHEPA,0702,027;%%
}

\lref\dimgiud{
 S.~Dimopoulos and G.~F.~Giudice,
  ``Multi-messenger theories of gauge-mediated supersymmetry breaking,''
  Phys.\ Lett.\  B {\bf 393}, 72 (1997)
  [arXiv:hep-ph/9609344].
  %%CITATION = PHLTA,B393,72;%%
}

\lref\pierre{S.~P.~Martin and P.~Ramond,
  ``Sparticle spectrum constraints,''
  Phys.\ Rev.\  D {\bf 48}, 5365 (1993)
  [arXiv:hep-ph/9306314].
  %%CITATION = PHRVA,D48,5365;%%
}

\lref\faraggi{
  A.~E.~Faraggi, J.~S.~Hagelin, S.~Kelley and D.~V.~Nanopoulos,
  ``Sparticle Spectroscopy,''
  Phys.\ Rev.\  D {\bf 45}, 3272 (1992).
  %%CITATION = PHRVA,D45,3272;%%
}

\lref\kawamura{
  Y.~Kawamura, H.~Murayama and M.~Yamaguchi,
  ``Probing symmetry breaking pattern using sfermion masses,''
  Phys.\ Lett.\  B {\bf 324}, 52 (1994)
  [arXiv:hep-ph/9402254].
  %%CITATION = PHLTA,B324,52;%%
}

%\MartinZB
\lref\MartinZB{
  S.~P.~Martin,
  ``Generalized messengers of supersymmetry breaking and the sparticle mass
  spectrum,''
  Phys.\ Rev.\  D {\bf 55}, 3177 (1997)
  [arXiv:hep-ph/9608224].
  %%CITATION = PHRVA,D55,3177;%%
}

\lref\spectroscopy{
 S.~Dimopoulos, S.~D.~Thomas and J.~D.~Wells,
  ``Sparticle spectroscopy and electroweak symmetry breaking with
  gauge-mediated supersymmetry breaking,''
  Nucl.\ Phys.\  B {\bf 488}, 39 (1997)
  [arXiv:hep-ph/9609434].
  %%CITATION = NUPHA,B488,39;%%
}

\lref\polchinski{ J.~Polchinski and L.~Susskind, ``Breaking Of
Supersymmetry At Intermediate-Energy,''
  Phys.\ Rev.\  D {\bf 26}, 3661 (1982).
  %%CITATION = PHRVA,D26,3661;%%
 }

\lref\dgp{ G.~R.~Dvali, G.~F.~Giudice and A.~Pomarol,
  ``The $\mu$-Problem in Theories with Gauge-Mediated Supersymmetry Breaking,''
  Nucl.\ Phys.\  B {\bf 478}, 31 (1996)
  [arXiv:hep-ph/9603238].
  %%CITATION = NUPHA,B478,31;%%
}

\lref\martinmu{ T.~S.~Roy and M.~Schmaltz,
  ``A hidden solution to the $\mu/B_\mu$ problem in gauge mediation,''
  arXiv:0708.3593 [hep-ph].
  %%CITATION = ARXIV:0708.3593;%%
}

\lref\hidrentwo{
  H.~Murayama, Y.~Nomura and D.~Poland,
  ``More Visible Effects of the Hidden Sector,''
  arXiv:0709.0775 [hep-ph].
  %%CITATION = ARXIV:0709.0775;%%
}

%\DineXI
\lref\DineXI{
  M.~Dine, N.~Seiberg and S.~Thomas,
  ``Higgs Physics as a Window Beyond the MSSM (BMSSM),''
  Phys.\ Rev.\  D {\bf 76}, 095004 (2007)
  [arXiv:0707.0005 [hep-ph]].
  %%CITATION = PHRVA,D76,095004;%%
}

%\WessCP
\lref\WessCP{
  J.~Wess and J.~Bagger,
  ``Supersymmetry and supergravity,''
%\href{http://www.slac.stanford.edu/spires/find/hep/www?irn=5426545}{SPIRES entry}
{\it  Princeton, USA: Univ. Pr. (1992) 259 p}
}

%\MartinNS
\lref\MartinNS{
  S.~P.~Martin,
  ``A supersymmetry primer,''
  arXiv:hep-ph/9709356.
  %%CITATION = HEP-PH/9709356;%%
}

%\KomargodskiAX
\lref\KomargodskiAX{
  Z.~Komargodski and N.~Seiberg,
  ``mu and General Gauge Mediation,''
  JHEP {\bf 0903}, 072 (2009)
  [arXiv:0812.3900 [hep-ph]].
  %%CITATION = JHEPA,0903,072;%%
}
%\MasonIQ
\lref\MasonIQ{
  J.~D.~Mason,
  ``Gauge Mediation with a small mu term and light squarks,''
  arXiv:0904.4485 [hep-ph].
  %%CITATION = ARXIV:0904.4485;%%
}

%\CsakiSR
\lref\CsakiSR{
  C.~Csaki, A.~Falkowski, Y.~Nomura and T.~Volansky,
  ``New Approach to the mu-Bmu Problem of Gauge-Mediated Supersymmetry
  Breaking,''
  Phys.\ Rev.\ Lett.\  {\bf 102}, 111801 (2009)
  [arXiv:0809.4492 [hep-ph]].
  %%CITATION = PRLTA,102,111801;%%
}

%\IntriligatorFR
\lref\IntriligatorFR{
  K.~A.~Intriligator and M.~Sudano,
  ``Comments on General Gauge Mediation,''
  JHEP {\bf 0811}, 008 (2008)
  [arXiv:0807.3942 [hep-ph]].
  %%CITATION = JHEPA,0811,008;%%
}

%\GorbatovQA
\lref\GorbatovQA{
  E.~Gorbatov and M.~Sudano,
  ``Sparticle Masses in Higgsed Gauge Mediation,''
  JHEP {\bf 0810}, 066 (2008)
  [arXiv:0802.0555 [hep-ph]].
  %%CITATION = JHEPA,0810,066;%%
}

%\LuoKF
\lref\LuoKF{
  M.~Luo and S.~Zheng,
  ``Gauge Extensions of Supersymmetric Models and Hidden Valleys,''
  JHEP {\bf 0904}, 122 (2009)
  [arXiv:0901.2613 [hep-ph]].
  %%CITATION = JHEPA,0904,122;%%
}

%\MeadeWD
\lref\MeadeWD{
  P.~Meade, N.~Seiberg and D.~Shih,
  ``General Gauge Mediation,''
  Prog.\ Theor.\ Phys.\ Suppl.\  {\bf 177}, 143 (2009)
  [arXiv:0801.3278 [hep-ph]].
  %%CITATION = PTPSA,177,143;%%
}

%\IntriligatorPY
\lref\IntriligatorPY{
  K.~A.~Intriligator, N.~Seiberg and D.~Shih,
  ``Supersymmetry Breaking, R-Symmetry Breaking and Metastable Vacua,''
  JHEP {\bf 0707}, 017 (2007)
  [arXiv:hep-th/0703281].
  %%CITATION = JHEPA,0707,017;%%
}

%\KomargodskiJF
\lref\KomargodskiJF{
  Z.~Komargodski and D.~Shih,
  ``Notes on SUSY and R-Symmetry Breaking in Wess-Zumino Models,''
  JHEP {\bf 0904}, 093 (2009)
  [arXiv:0902.0030 [hep-th]].
  %%CITATION = JHEPA,0904,093;%%
}

%\ShiraiRR
\lref\ShiraiRR{
  S.~Shirai, M.~Yamazaki and K.~Yonekura,
  ``Aspects of Non-minimal Gauge Mediation,''
  JHEP {\bf 1006}, 056 (2010)
  [arXiv:1003.3155 [hep-ph]].
  %%CITATION = JHEPA,1006,056;%%
}

%\SeibergQJ
\lref\SeibergQJ{
  N.~Seiberg, T.~Volansky and B.~Wecht,
  ``Semi-direct Gauge Mediation,''
  JHEP {\bf 0811}, 004 (2008)
  [arXiv:0809.4437 [hep-ph]].
  %%CITATION = JHEPA,0811,004;%%
}

%\ElvangGK
\lref\ElvangGK{
  H.~Elvang and B.~Wecht,
  ``Semi-Direct Gauge Mediation with the 4-1 Model,''
  JHEP {\bf 0906}, 026 (2009)
  [arXiv:0904.4431 [hep-ph]].
  %%CITATION = JHEPA,0906,026;%%
}

%\ChengAN
\lref\ChengAN{
  H.~C.~Cheng, D.~E.~Kaplan, M.~Schmaltz and W.~Skiba,
  ``Deconstructing gaugino mediation,''
  Phys.\ Lett.\  B {\bf 515}, 395 (2001)
  [arXiv:hep-ph/0106098].
  %%CITATION = PHLTA,B515,395;%%
}

%\CsakiEM
\lref\CsakiEM{
  C.~Csaki, J.~Erlich, C.~Grojean and G.~D.~Kribs,
  ``4D constructions of supersymmetric extra dimensions and gaugino
  mediation,''
  Phys.\ Rev.\  D {\bf 65}, 015003 (2002)
  [arXiv:hep-ph/0106044].
  %%CITATION = PHRVA,D65,015003;%%
}

%\DumitrescuHA
\lref\DumitrescuHA{
  T.~T.~Dumitrescu, Z.~Komargodski, N.~Seiberg and D.~Shih,
  ``General Messenger Gauge Mediation,''
  JHEP {\bf 1005}, 096 (2010)
  [arXiv:1003.2661 [hep-ph]].
  %%CITATION = JHEPA,1005,096;%%
}

%\FrancoWF
\lref\FrancoWF{
  S.~Franco and S.~Kachru,
  ``Single-Sector Supersymmetry Breaking in Supersymmetric QCD,''
  Phys.\ Rev.\  D {\bf 81}, 095020 (2010)
  [arXiv:0907.2689 [hep-th]].
  %%CITATION = PHRVA,D81,095020;%%}
}

%\IntriligatorDD
\lref\IntriligatorDD{
  K.~A.~Intriligator, N.~Seiberg and D.~Shih,
  ``Dynamical SUSY breaking in meta-stable vacua,''
  JHEP {\bf 0604}, 021 (2006)
  [arXiv:hep-th/0602239].
  %%CITATION = JHEPA,0604,021;%%
}

%\SeibergPQ
\lref\SeibergPQ{
  N.~Seiberg,
  ``Electric - magnetic duality in supersymmetric nonAbelian gauge theories,''
  Nucl.\ Phys.\  B {\bf 435}, 129 (1995)
  [arXiv:hep-th/9411149].
  %%CITATION = NUPHA,B435,129;%%
}

%\GiveonWP
\lref\GiveonWP{
  A.~Giveon, A.~Katz and Z.~Komargodski,
  ``On SQCD with massive and massless flavors,''
  JHEP {\bf 0806}, 003 (2008)
  [arXiv:0804.1805 [hep-th]].
  %%CITATION = JHEPA,0806,003;%%
}

%\IntriligatorID
\lref\IntriligatorID{
  K.~A.~Intriligator and N.~Seiberg,
  ``Duality, monopoles, dyons, confinement and oblique confinement in
  supersymmetric SO(N(c)) gauge theories,''
  Nucl.\ Phys.\  B {\bf 444}, 125 (1995)
  [arXiv:hep-th/9503179].
  %%CITATION = NUPHA,B444,125;%%
}

%\IntriligatorNE
\lref\IntriligatorNE{
  K.~A.~Intriligator and P.~Pouliot,
  ``Exact superpotentials, quantum vacua and duality in supersymmetric SP(N(c))
  gauge theories,''
  Phys.\ Lett.\  B {\bf 353}, 471 (1995)
  [arXiv:hep-th/9505006].
  %%CITATION = PHLTA,B353,471;%%
}

%\KaplanAC
\lref\KaplanAC{
  D.~E.~Kaplan, G.~D.~Kribs and M.~Schmaltz,
  ``Supersymmetry breaking through transparent extra dimensions,''
  Phys.\ Rev.\  D {\bf 62}, 035010 (2000)
  [arXiv:hep-ph/9911293].
  %%CITATION = PHRVA,D62,035010;%%
}

%\ChackoMI
\lref\ChackoMI{
  Z.~Chacko, M.~A.~Luty, A.~E.~Nelson and E.~Ponton,
  ``Gaugino mediated supersymmetry breaking,''
  JHEP {\bf 0001}, 003 (2000)
  [arXiv:hep-ph/9911323].
  %%CITATION = JHEPA,0001,003;%%
}

%\EllisKD
\lref\EllisKD{
  J.~R.~Ellis, K.~Enqvist and D.~V.~Nanopoulos,
  ``A Very Light Gravitino In A No Scale Model,''
  Phys.\ Lett.\  B {\bf 147}, 99 (1984).
  %%CITATION = PHLTA,B147,99;%%
}

%\EllisBM
\lref\EllisBM{
  J.~R.~Ellis, C.~Kounnas and D.~V.~Nanopoulos,
  ``No Scale Supersymmetric Guts,''
  Nucl.\ Phys.\  B {\bf 247}, 373 (1984).
  %%CITATION = NUPHA,B247,373;%%
}

%\MurayamaGE
\lref\MurayamaGE{
  H.~Murayama, Y.~Nomura and D.~Poland,
  ``More Visible Effects of the Hidden Sector,''
  Phys.\ Rev.\  D {\bf 77}, 015005 (2008)
  [arXiv:0709.0775 [hep-ph]].
  %%CITATION = PHRVA,D77,015005;%%
}

%\RoyNZ
\lref\RoyNZ{
  T.~S.~Roy and M.~Schmaltz,
  ``A hidden solution to the $\mu/B_{\mu}$ problem in gauge mediation,''
  Phys.\ Rev.\  D {\bf 77}, 095008 (2008)
  [arXiv:0708.3593 [hep-ph]].
  %%CITATION = PHRVA,D77,095008;%%
}

%\GiveonYU
\lref\GiveonYU{
  A.~Giveon, A.~Katz and Z.~Komargodski,
  ``Uplifted Metastable Vacua and Gauge Mediation in SQCD,''
  JHEP {\bf 0907}, 099 (2009)
  [arXiv:0905.3387 [hep-th]].
  %%CITATION = JHEPA,0907,099;%%
}

%\AuzziWM
\lref\AuzziWM{
  R.~Auzzi, S.~Elitzur and A.~Giveon,
  ``On Uplifted SUSY-Breaking Vacua and Direct Mediation in Generalized SQCD,''
  JHEP {\bf 1003}, 094 (2010)
  [arXiv:1001.1234 [hep-th]].
  %%CITATION = JHEPA,1003,094;%%
}

%\KitanoFA
\lref\KitanoFA{
  R.~Kitano, H.~Ooguri and Y.~Ookouchi,
 ``Supersymmetry Breaking and Gauge Mediation,''
  arXiv:1001.4535 [hep-th].
  %%CITATION = ARXIV:1001.4535;%%
}

%\MeadeWD
\lref\MeadeWD{
  P.~Meade, N.~Seiberg and D.~Shih,
  ``General Gauge Mediation,''
  Prog.\ Theor.\ Phys.\ Suppl.\  {\bf 177}, 143 (2009)
  [arXiv:0801.3278 [hep-ph]].
  %%CITATION = PTPSA,177,143;%%
}

%\IntriligatorFE
\lref\IntriligatorFE{
  K.~Intriligator, D.~Shih and M.~Sudano,
  ``Surveying Pseudomoduli: the Good, the Bad and the Incalculable,''
  JHEP {\bf 0903}, 106 (2009)
  [arXiv:0809.3981 [hep-th]].
  %%CITATION = JHEPA,0903,106;%%
}

%\BarnesJJ
\lref\BarnesJJ{
  E.~Barnes, K.~A.~Intriligator, B.~Wecht and J.~Wright,
  ``Evidence for the strongest version of the 4d a-theorem, via  a-maximization
  along RG flows,''
  Nucl.\ Phys.\  B {\bf 702}, 131 (2004)
  [arXiv:hep-th/0408156].
  %%CITATION = NUPHA,B702,131;%%
}

%\AmaritiSZ
\lref\AmaritiSZ{
 A.~Amariti, L.~Girardello, A.~Mariotti and M.~Siani,
 ``Metastable Vacua in Superconformal SQCD-like Theories,''
 arXiv:1003.0523 [hep-th].
 %%CITATION = ARXIV:1003.0523;%%
}

%\EssigKZ
\lref\EssigKZ{
  R.~Essig, J.~F.~Fortin, K.~Sinha, G.~Torroba and M.~J.~Strassler,
  ``Metastable supersymmetry breaking and multitrace deformations of SQCD,''
  JHEP {\bf 0903}, 043 (2009)
  [arXiv:0812.3213 [hep-th]].
  %%CITATION = JHEPA,0903,043;%%
}

%\GiveonNE
\lref\GiveonNE{
  A.~Giveon, A.~Katz, Z.~Komargodski and D.~Shih,
  ``Dynamical SUSY and R-symmetry breaking in SQCD with massive and massless
  flavors,''
  JHEP {\bf 0810}, 092 (2008)
  [arXiv:0808.2901 [hep-th]].
  %%CITATION = JHEPA,0810,092;%%
}

%\AbelJX
\lref\AbelJX{
  S.~Abel, C.~Durnford, J.~Jaeckel and V.~V.~Khoze,
  ``Dynamical breaking of $U(1)_{R}$ and supersymmetry in a metastable vacuum,''
  Phys.\ Lett.\  B {\bf 661}, 201 (2008)
  [arXiv:0707.2958 [hep-ph]].
  %%CITATION = PHLTA,B661,201;%%
}

%\DineXT
\lref\DineXT{
  M.~Dine and J.~Mason,
  ``Gauge mediation in metastable vacua,''
  Phys.\ Rev.\  D {\bf 77}, 016005 (2008)
  [arXiv:hep-ph/0611312].
  %%CITATION = PHRVA,D77,016005;%%
}

%\KitanoXG
\lref\KitanoXG{
  R.~Kitano, H.~Ooguri and Y.~Ookouchi,
  ``Direct mediation of meta-stable supersymmetry breaking,''
  Phys.\ Rev.\  D {\bf 75}, 045022 (2007)
  [arXiv:hep-ph/0612139].
  %%CITATION = PHRVA,D75,045022;%%
}

%\DeSimoneWS
\lref\DeSimoneWS{
  A.~De Simone, J.~Fan, V.~Sanz and W.~Skiba,
  ``Leptogenic Supersymmetry,''
  Phys.\ Rev.\  D {\bf 80}, 035010 (2009)
  [arXiv:0903.5305 [hep-ph]].
  %%CITATION = PHRVA,D80,035010;%%
}

%\DineGM
\lref\DineGM{
  M.~Dine, J.~L.~Feng and E.~Silverstein,
  ``Retrofitting O'Raifeartaigh models with dynamical scales,''
  Phys.\ Rev.\  D {\bf 74}, 095012 (2006)
  [arXiv:hep-th/0608159].
  %%CITATION = PHRVA,D74,095012;%%
}

%\KatzGH
\lref\KatzGH{
  A.~Katz,
  ``On the Thermal History of Calculable Gauge Mediation,''
  JHEP {\bf 0910}, 054 (2009)
  [arXiv:0907.3930 [hep-th]].
  %%CITATION = JHEPA,0910,054;%%
}

%\DeSimoneGM
\lref\DeSimoneGM{
  A.~De Simone, J.~Fan, M.~Schmaltz and W.~Skiba,
  ``Low-scale gaugino mediation, lots of leptons at the LHC,''
  Phys.\ Rev.\  D {\bf 78}, 095010 (2008)
  [arXiv:0808.2052 [hep-ph]].
  %%CITATION = PHRVA,D78,095010;%%
}

%\GiveonEF
\lref\GiveonEF{
  A.~Giveon and D.~Kutasov,
  ``Stable and Metastable Vacua in SQCD,''
  Nucl.\ Phys.\  B {\bf 796}, 25 (2008)
  [arXiv:0710.0894 [hep-th]].
  %%CITATION = NUPHA,B796,25;%%
}

%\GiveonEW
\lref\GiveonEW{
  A.~Giveon and D.~Kutasov,
  ``Stable and Metastable Vacua in Brane Constructions of SQCD,''
  JHEP {\bf 0802}, 038 (2008)
  [arXiv:0710.1833 [hep-th]].
  %%CITATION = JHEPA,0802,038;%%
}

%\KaplanAV
\lref\KaplanAV{
  D.~E.~Kaplan and T.~M.~P.~Tait,
  ``Supersymmetry breaking, fermion masses and a small extra dimension,''
  JHEP {\bf 0006}, 020 (2000)
  [arXiv:hep-ph/0004200].
  %%CITATION = JHEPA,0006,020;%%
}

%\KobayashiRN
\lref\KobayashiRN{
  T.~Kobayashi, Y.~Nakai and R.~Takahashi,
  ``Fine Tuning in General Gauge Mediation,''
  JHEP {\bf 1001}, 003 (2010)
  [arXiv:0910.3477 [hep-ph]].
  %%CITATION = JHEPA,1001,003;%%
}

%\AbelVE
\lref\AbelVE{
  S.~Abel, M.~J.~Dolan, J.~Jaeckel and V.~V.~Khoze,
  ``Phenomenology of Pure General Gauge Mediation,''
  JHEP {\bf 0912}, 001 (2009)
  [arXiv:0910.2674 [hep-ph]].
  %%CITATION = JHEPA,0912,001;%%
}

%\AbelZE
\lref\AbelZE{
  S.~A.~Abel, J.~Jaeckel and V.~V.~Khoze,
  ``Gaugino versus Sfermion Masses in Gauge Mediation,''
  Phys.\ Lett.\  B {\bf 682}, 441 (2010)
  [arXiv:0907.0658 [hep-ph]].
  %%CITATION = PHLTA,B682,441;%%
}

%\RajaramanGA
\lref\RajaramanGA{
  A.~Rajaraman, Y.~Shirman, J.~Smidt and F.~Yu,
  ``Parameter Space of General Gauge Mediation,''
  Phys.\ Lett.\  B {\bf 678}, 367 (2009)
  [arXiv:0903.0668 [hep-ph]].
  %%CITATION = PHLTA,B678,367;%%
}

%\LuoKF
\lref\LuoKF{
  M.~Luo and S.~Zheng,
  ``Gauge Extensions of Supersymmetric Models and Hidden Valleys,''
  JHEP {\bf 0904}, 122 (2009)
  [arXiv:0901.2613 [hep-ph]].
  %%CITATION = JHEPA,0904,122;%%
}

%\MarquesYU
\lref\MarquesYU{
  D.~Marques,
  ``Generalized messenger sector for gauge mediation of supersymmetry breaking
  and the soft spectrum,''
  JHEP {\bf 0903}, 038 (2009)
  [arXiv:0901.1326 [hep-ph]].
  %%CITATION = JHEPA,0903,038;%%
}

%\CarpenterHE
\lref\CarpenterHE{
  L.~M.~Carpenter,
  ``Surveying the Phenomenology of General Gauge Mediation,''
  arXiv:0812.2051 [hep-ph].
  %%CITATION = ARXIV:0812.2051;%%
}

%\BenakliPG
\lref\BenakliPG{
  K.~Benakli and M.~D.~Goodsell,
  ``Dirac Gauginos in General Gauge Mediation,''
  Nucl.\ Phys.\  B {\bf 816}, 185 (2009)
  [arXiv:0811.4409 [hep-ph]].
  %%CITATION = NUPHA,B816,185;%%
}

%\MarquesVA
\lref\MarquesVA{
  D.~Marques and F.~A.~Schaposnik,
  ``Explicit R-Symmetry Breaking and Metastable Vacua,''
  JHEP {\bf 0811}, 077 (2008)
  [arXiv:0809.4618 [hep-th]].
  %%CITATION = JHEPA,0811,077;%%
}

%\CarpenterWI
\lref\CarpenterWI{
  L.~M.~Carpenter, M.~Dine, G.~Festuccia and J.~D.~Mason,
  ``Implementing General Gauge Mediation,''
  Phys.\ Rev.\  D {\bf 79}, 035002 (2009)
  [arXiv:0805.2944 [hep-ph]].
  %%CITATION = PHRVA,D79,035002;%%
}

%\CheungES
\lref\CheungES{
  C.~Cheung, A.~L.~Fitzpatrick and D.~Shih,
  ``(Extra)Ordinary Gauge Mediation,''
  JHEP {\bf 0807}, 054 (2008)
  [arXiv:0710.3585 [hep-ph]].
  %%CITATION = JHEPA,0807,054;%%
}

%\ArkaniHamedNC
\lref\ArkaniHamedNC{
  N.~Arkani-Hamed, A.~G.~Cohen and H.~Georgi,
  ``Electroweak symmetry breaking from dimensional deconstruction,''
  Phys.\ Lett.\  B {\bf 513}, 232 (2001)
  [arXiv:hep-ph/0105239].
  %%CITATION = PHLTA,B513,232;%%
}

%\KatzXG
\lref\KatzXG{
  A.~Katz and B.~Tweedie,
  ``Leptophilic Signals of a Sneutrino (N)LSP and Flavor Biases from
  Flavor-Blind SUSY,''
  Phys.\ Rev.\  D {\bf 81}, 115003 (2010)
  [arXiv:1003.5664 [hep-ph]].
  %%CITATION = PHRVA,D81,115003;%%
}

%\KatzQX
\lref\KatzQX{
  A.~Katz and B.~Tweedie,
  ``Signals of a Sneutrino (N)LSP at the LHC,''
  Phys.\ Rev.\  D {\bf 81}, 035012 (2010)
  [arXiv:0911.4132 [hep-ph]].
  %%CITATION = PHRVA,D81,035012;%%
}

%\MeadeQV
\lref\MeadeQV{
  P.~Meade, M.~Reece and D.~Shih,
  ``Prompt Decays of General Neutralino NLSPs at the Tevatron,''
  JHEP {\bf 1005}, 105 (2010)
  [arXiv:0911.4130 [hep-ph]].
  %%CITATION = JHEPA,1005,105;%%
}

%\SchaferNamekiIZ
\lref\SchaferNamekiIZ{
  S.~Schafer-Nameki, C.~Tamarit and G.~Torroba,
  ``A Hybrid Higgs,''
  arXiv:1005.0841 [hep-ph].
  %%CITATION = ARXIV:1005.0841;%%
}

%\BuicanWS
\lref\BuicanWS{
  M.~Buican, P.~Meade, N.~Seiberg and D.~Shih,
  ``Exploring General Gauge Mediation,''
  JHEP {\bf 0903}, 016 (2009)
  [arXiv:0812.3668 [hep-ph]].
  %%CITATION = JHEPA,0903,016;%%
}

%\CraigHF
\lref\CraigHF{
  N.~Craig, R.~Essig, S.~Franco, S.~Kachru and G.~Torroba,
  ``Dynamical Supersymmetry Breaking, with Flavor,''
  Phys.\ Rev.\  D {\bf 81}, 075015 (2010)
  [arXiv:0911.2467 [hep-ph]].
  %%CITATION = PHRVA,D81,075015;%%
}

%\NelsonNF
\lref\NelsonNF{
  A.~E.~Nelson and N.~Seiberg,
  ``R symmetry breaking versus supersymmetry breaking,''
  Nucl.\ Phys.\  B {\bf 416}, 46 (1994)
  [arXiv:hep-ph/9309299].
  %%CITATION = NUPHA,B416,46;%%
}

%\IntriligatorPY
\lref\IntriligatorPY{
  K.~A.~Intriligator, N.~Seiberg and D.~Shih,
  ``Supersymmetry Breaking, R-Symmetry Breaking and Metastable Vacua,''
  JHEP {\bf 0707}, 017 (2007)
  [arXiv:hep-th/0703281].
  %%CITATION = JHEPA,0707,017;%%
}

%\IntriligatorCP
\lref\IntriligatorCP{
  K.~A.~Intriligator and N.~Seiberg,
  ``Lectures on Supersymmetry Breaking,''
  Class.\ Quant.\ Grav.\  {\bf 24}, S741 (2007)
  [arXiv:hep-ph/0702069].
  %%CITATION = CQGRD,24,S741;%%
}

%\GiudiceBP
\lref\GiudiceBP{
  G.~F.~Giudice and R.~Rattazzi,
  ``Theories with gauge-mediated supersymmetry breaking,''
  Phys.\ Rept.\  {\bf 322}, 419 (1999)
  [arXiv:hep-ph/9801271].
  %%CITATION = PRPLC,322,419;%%
}

%\HillMU
\lref\HillMU{
  C.~T.~Hill, S.~Pokorski and J.~Wang,
  ``Gauge invariant effective Lagrangian for Kaluza-Klein modes,''
  Phys.\ Rev.\  D {\bf 64}, 105005 (2001)
  [arXiv:hep-th/0104035].
  %%CITATION = PHRVA,D64,105005;%%
}

%\ArkaniHamedCA
\lref\ArkaniHamedCA{
  N.~Arkani-Hamed, A.~G.~Cohen and H.~Georgi,
  ``(De)constructing dimensions,''
  Phys.\ Rev.\ Lett.\  {\bf 86}, 4757 (2001)
  [arXiv:hep-th/0104005].
  %%CITATION = PRLTA,86,4757;%%
}

%\HabaRJ
\lref\HabaRJ{
  N.~Haba and N.~Maru,
  ``A Simple Model of Direct Gauge Mediation of Metastable Supersymmetry
  Breaking,''
  Phys.\ Rev.\  D {\bf 76}, 115019 (2007)
  [arXiv:0709.2945 [hep-ph]].
  %%CITATION = PHRVA,D76,115019;%%
}

%\MaruYX
\lref\MaruYX{
  N.~Maru,
  ``Direct Gauge Mediation of Uplifted Metastable Supersymmetry Breaking in
  Supergravity,''
  arXiv:1008.1440 [hep-ph].
  %%CITATION = ARXIV:1008.1440;%%
}

%\McGarrieKH
\lref\McGarrieKH{
  M.~McGarrie and R.~Russo,
  ``General Gauge Mediation in 5D,''
  Phys.\ Rev.\  D {\bf 82}, 035001 (2010)
  [arXiv:1004.3305 [hep-ph]].
  %%CITATION = PHRVA,D82,035001;%%
}

\Title{\vbox{\baselineskip12pt \hbox{UMD-PP-10-013}}}
{\vbox{\centerline{Direct Gaugino Mediation} }}
\medskip

\centerline{ Daniel Green,$^{\spadesuit}$ Andrey
Katz,$^{\clubsuit}$ and Zohar Komargodski$^{\spadesuit}$}

\bigskip
\centerline{\it $^\spadesuit$School of Natural Sciences, Institute for
Advanced Study,  Princeton, NJ 08540}
\centerline{\it $^\clubsuit$Department of Physics, University of Maryland, College Park, MD 20742 }

\smallskip

\vglue .3cm

\bigskip
\noindent

We describe renormalizable supersymmetric four-dimensional
theories which lead to gaugino mediation and various
generalizations thereof. Even though these models are strongly
coupled, we can demonstrate the parametric suppression of soft
scalar masses via Seiberg duality. For instance, we show that our
models have a parameter which continuously interpolates between
suppressed soft scalar masses and their conventional gauge
mediated contribution. The main physical effect which we utilize
is the general relation between massive deformations in one frame
and higgsing in the dual frame. Some compelling and relatively
unexplored particle physics scenarios arise naturally in this
framework. We offer preliminary comments on various aspects of the
phenomenology and outline several of the outstanding open
problems.

\Date{August/2010}

%\draftmode

%%%%%%%%%%%%%%%%%%%%%%%%

\newsec{Introduction}

Weak scale supersymmetry (SUSY) is an attractive framework for
addressing the hierarchy problem.  Within the context of the
Minimally Supersymmetric Standard Model (MSSM), the Higgs sector
can be made natural provided the soft masses for the gauginos and
stops are both comparable to the weak scale.  Achieving such a
spectrum while solving the flavor problem has long been a
benchmark for successful model building.  Gauge
mediation~\refs{\DineGU\DimopoulosAU\NappiHM\DineAG\AlvarezGaumeWY\DineZB\DineVC-\DineYW}
is a promising scenario as it solves the flavor problem by
construction.  Furthermore, the simplest schemes for gauge mediation,
like minimal gauge mediation~\refs{\DineZB\DineVC-\DineYW}, also
provide a desirable spectrum of soft masses.

Since the work of ISS~\refs{\IntriligatorDD}, there has been a
rejuvenated interest in embedding gauge mediation in models of
dynamical SUSY breaking (for many relevant references see the
recent review~\KitanoFA). However, in most cases the sparticles
turned out to be heavy compared to the gauginos. This is
associated to the need to break $R$-symmetry in order to generate
gaugino masses, and in addition, to constraints on the vacuum
structure of the theory~\refs{\KomargodskiJF,\ShiraiRR}.

Gaugino mediation~\refs{\KaplanAC,\ChackoMI} is a different scheme
of mediation that gives a spectrum where the scalar masses are no
larger than the gaugino masses. Gaugino mediation arises when
there is some mechanism that screens the scalar masses from SUSY
breaking without altering the gaugino masses.  As a result, the
scalar masses are smaller than the gaugino masses at high energies
but are generated by renormalization group flow in the MSSM. There
are several proposals for how to achieve such screening, starting
with no-scale supergravity~\refs{\EllisKD,\EllisBM} and including
large extra dimensions~\refs{\KaplanAC,\ChackoMI},
deconstruction~\refs{\ArkaniHamedCA\HillMU\ChengAN-\CsakiEM},
large anomalous dimensions~\refs{\RoyNZ,\MurayamaGE}, or large
numbers of messengers. However, some of these scenarios are not on
theoretically firm grounds; for example, some run into strong
coupling and are, strictly speaking, incalculable, while others
lack consistent examples or UV completions.

In this paper we construct renormalizable four-dimensional
dynamical models of SUSY breaking that lead to scalar screening.
We will also show that such models are generic. The existence of a
calculable framework allows detailed phenomenological studies, and
indeed, models of scalar screening can lead to novel,
well-motivated, particle physics scenarios. At low energies, the
effective Lagrangians of our models describe a sequence of
symmetry breaking phenomena. The structure of the effective theory
has some features in common with the ansatz arising via
deconstructing higher dimensional
models~\refs{\ArkaniHamedCA\HillMU\ChengAN-\CsakiEM}. The
suppression of the scalar masses is given either by the ratio of
some higgsing scale $v$ and the messenger mass $M$, or additional
loop factors.

Of course, our results differ from simple deconstructed models.
Their ad-hoc structure of higgsing and ``link
fields'' arises in our case dynamically from a single sector
theory, which undergoes strong dynamics along its flow. The
relation between models inspired by the ansatz of deconstruction
and our dynamical models is analogous to the relation between
O'Raifeartaigh-like models and calculable models of dynamical SUSY
breaking: calculable models of dynamical SUSY breaking often
reduce to O'R-like models in the IR. Indeed, similarly to O'R-like
models, deconstructed models leave many conceptual questions open.
Finding elegant dynamical models which flow to them provides some
answers. On the other hand, this analogy also suggests of the
universality of deconstructed models and their importance,
something we believe in strongly.

Even though we discuss strongly coupled theories, we are able to
analyze them thanks to Seiberg duality~\SeibergPQ. A well known
phenomenon in electric-magnetic duality is that a mass term in one
duality frame is described by higgsing in the other duality frame.
As we will explain, certain patterns of higgsing are sufficient
for generating suppression factors in the soft scalar masses. This
suggests that gaugino mediation may arise from mass deformations
of asymptotically free theories with IR free magnetic dual
descriptions. Luckily, the same kind of deformations lead to
metastable SUSY breaking~\IntriligatorDD\ and so we claim that
theories with suppressed scalar masses are generic in this class.

Strictly speaking, our models fall into the class of General Gauge
Mediation (GGM) as defined in~\MeadeWD, so they are really
dynamical models of gauge mediation. However, they can yield
spectra far from those we are used to. We have therefore chosen to
refer to them as ``Direct Gaugino Mediation'' models.\foot{The
analysis of~\MeadeWD\ has stimulated various attempts to
generalize the minimal setup of gauge mediation. A partial list of
such works is~\refs{\CarpenterWI\BenakliPG\CarpenterHE\BuicanWS
\MarquesYU\LuoKF\RajaramanGA\AbelVE\KobayashiRN-\McGarrieKH}.}

The plan for the paper is as follows.  In section~2, we  explain
the low energy theory of scalar screening in detail, including how
to understand it in the context of general gauge mediation. We
will see that the standard model charges of the messengers are
effectively functions of energy.  In section~3, we will present
detailed models based on deformations of SQCD with $SU(N)$, and
further generalize to $SO(N)$ and $Sp(N)$ gauge groups.  In
section~4, we will include more phenomenologically relevant
details of the model and many model building applications. We will
speculate about the kind of spectra such models can lead to, and
discuss preliminary aspects of the phenomenology. Section~5
contains a summary and an outlook.

%%%%%%%%%%%%%%%%%%%%%%%%%%%%%%%%%%%%%%%%%%%%%%%%%%%%%%%%%%%%%%%
%%%%%%%%%%%%%%%%%%%%%%%%%%%%%%%%%%%%%%%%%%%%%%%%%%%%%%%%%%%%%%%

\newsec{From Gauge to Gaugino Mediation}

We begin this section with a review of some known facts about
gauge mediation and introduce the question of gaugino mediation.
This is also our opportunity to explain in more detail what we are
after in this paper. On occasion, we use the language of GGM to
shed some further light on the problem.

Gauge mediation is a compelling way of mediating SUSY breaking in
some hidden sector $H$ to the MSSM. We assume that the flavor
symmetry of $H$, $F_H$, contains the MSSM group $G_{MSSM}\subset
F_H$ and we gauge this subgroup. If the typical mass of particles
in $H$ is $M$ and the SUSY breaking scale is $F$ then we expect
that this mechanism generates soft scalar and gaugino masses of
the form \eqn\softmasses{m_{\tilde f}^2\sim {\alpha^2\over
(4\pi)^2}{F^2\over M^2} ~,\qquad m_{\tilde g}\sim {\alpha\over
4\pi}{F\over M}~.} This has all the usual nice features of gauge
mediation, especially that it is flavor blind and that the gaugino
and scalar soft masses are comparable.

We can test~\softmasses\ in many of the calculable models of SUSY
breaking. It is rather easy to engineer models with messengers
which lead to~\softmasses, but the success with dynamical models
has been limited. In many of these models the scalar masses are
too large compared to the gaugino masses (unless cumbersome tricks
are employed, e.g.~\refs{\KitanoXG\GiveonYU\AuzziWM-\MaruYX}) and so the
resulting spectrum is fine tuned.

In this paper we would like to look for models which lead
dynamically to boundary conditions with $m_{\tilde g}\gg m_{\tilde
f}$. In light of the description of the situation with gauge
mediation above, this goal might seem hopeless, but we will see
that in fact the opposite is true. Let us describe an {\it ansatz}
for how this can be achieved in
principle~\refs{\ChengAN,\CsakiEM}.

Consider a copy of the SM group $G_{SM}^{(2)}$ embedded in the
flavor group $F_H$ of some SUSY-breaking sector (with typical
scale $M$ and SUSY breaking scale $F$). Then we have some sector
of ``link fields'' which are charged under both $G_{SM}^{(2)}$ and
another copy of the SM group, $G_{SM}^{(1)}$. The sector of link
fields is assumed to be supersymmetric when we turn off the gauge
coupling of $G_{SM}^{(2)}$. An important dynamical assumption is
that the sector of link fields breaks the symmetry
\eqn\symmbre{G_{SM}^{(1)}\times G_{SM}^{(2)}\hookrightarrow
G_{SM}~,} at some scale $v$. This symmetry breaking should take
place in the supersymmetric limit, when the gauge coupling of
$G_{SM}^{(2)}$ is zero. The observable gauge fields are identified
with the unbroken ones, $G_{SM}$. All the MSSM matter fields are
assumed to be charged only under $G_{SM}^{(1)}$. The situation is
summarized in Fig.1.

\bigskip
\bigskip
\centerline{\epsfxsize=0.55\hsize\epsfbox{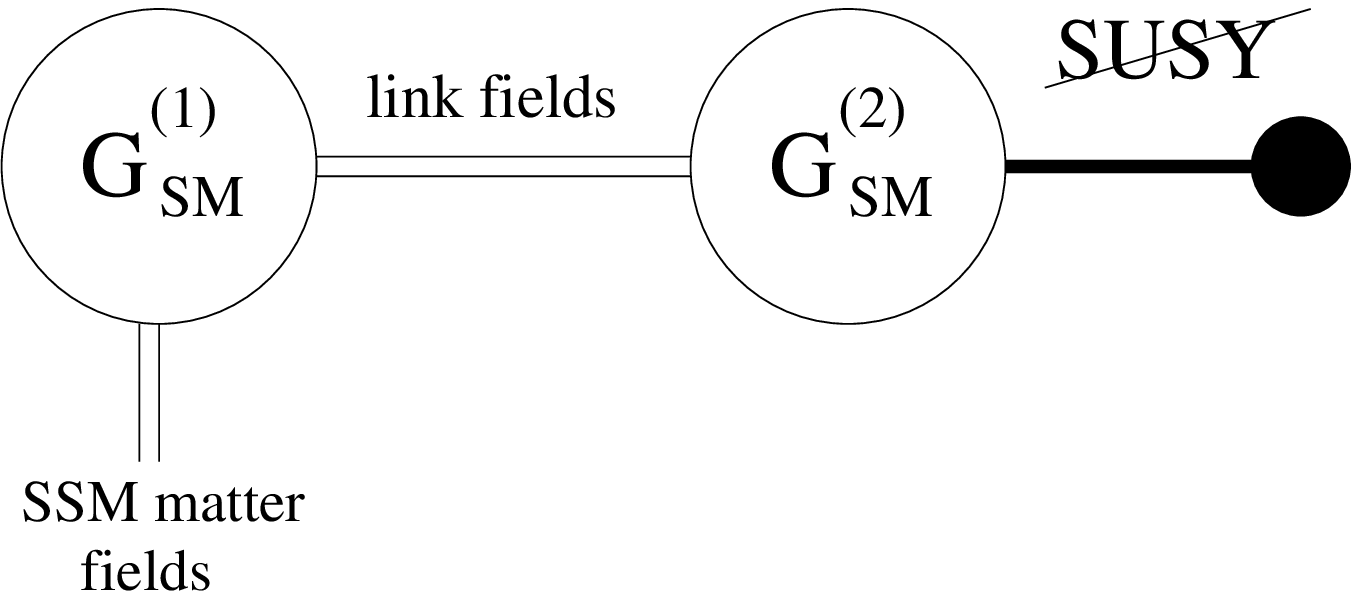}}
\noindent{\ninepoint\sl \baselineskip=8pt {\bf Figure 1}:{\sl $\;$
A schematic picture of the deconstruction ansatz for gaugino
mediation. }}
\bigskip

To understand what these models do, we first assume that $v\gg M$.
Below the scale $v$ the theory is still supersymmetric, including
the hidden sector. Both the hidden sector and the visible sector
are charged under $G_{SM}$. Thus, it looks precisely like the
usual setup of gauge mediation with $G_{SM}$ being embedded in the
flavor group $F_H$. We recover the conventional paradigm and
predictions. Now let us assume that $v\ll M$. This limit is much
more interesting. Here, at energies above $v$ the visible matter
fields do not interact with the SUSY breaking sector up to
diagrams with four loops in the various gauge couplings. So the
contribution from these energy scales is negligible and can be
estimated as $m_{\tilde f}^2\sim {\alpha^4\over (4\pi)^4}{F^2\over
M^2}$. At energies below $v$ the group breaks~\symmbre\ and now
there are interactions at two loops and three loops between the
sparticles of the MSSM and the SUSY breaking sector. Since $v\ll
M$ these two loop interactions cannot probe energy scales of order
$M$ and the scalar masses come out suppressed $m_{\tilde f}^2\sim
{\alpha^2\over (4\pi)^2}{F^2\over M^2}{v^2\over M^2}$. Note the
$v^2/M^2$ suppression compared to~\softmasses. Some three loop
interactions are important as well, but we will postpone this
topic to section~4.

To summarize, these models with $v\ll M$ give rise to scalar
masses which are parametrically smaller than the conventional
estimate.  Their actual value at the boundary is determined by
whether the $v^2/M^2$ suppression is larger or smaller than the
additional loop factor suppression.

The gaugino soft mass behaves very differently (and much simpler).
As far as the the gauge group $G_{SM}^{(2)}$ is concerned, we
generate, via the usual gauge mediation mechanism, gaugino mass of
the form $m_{\tilde g^{(2)}}\sim {\alpha\over 4\pi}{F\over M}$.
(Of course in this case, by $\alpha$ we mean $\alpha_2$, but we
prefer not to clutter the notation at this stage.) The breaking to
the diagonal implies that the actual visible gaugino is a linear
combination of $\tilde g^{(1)}$ and $\tilde g^{(2)}$. This means
that it directly acquires the soft mass of $\tilde g^{(2)}$. This
soft mass is, consequently, {\it independent} of the higgsing
scale $v$. In other words, the soft mass of the physical low
energy gaugino is $m_{\tilde g}\sim {\alpha\over 4\pi}{F\over M}$.

We see that the ansatz of Fig.1 with $v\ll M$ gives rise to
boundary conditions in which the gauginos are parametrically
heavier than the sparticles.

For an alternate explanation of why the gaugino is unsuppressed (and
independent of $v$) while the scalars are, it is useful to invoke
the language of GGM. If we think about the link fields together
with the SUSY breaking sector $H$ and the gauge fields
$G_{SM}^{(2)}$ as one hidden sector, then this is the setup of
GGM. We can write the scalar masses in terms of correlation
functions of $G_{SM}^{(1)}$ global currents as\foot{We are using
the notation of~\MeadeWD.} \eqn\scalarsGGM{m_{\tilde f}^2\sim
g_{1}^4\int {d^4p\over
p^2}\left(C_{0}^{(1)}(p^2)-4C_{1/2}^{(1)}(p^2)+3C_{1}^{(1)}(p^2)\right)~.}
The superscript is to remind that these are correlation functions
associated to the global symmetry $G_{SM}^{(1)}$. The gaugino mass
is given by \eqn\gauginoGGM{m_{\tilde g }\sim
g_{1}^2B_{1/2}(p=0)~.}

The key difference between the scalar and gaugino mass is that the
former is sensitive to correlation functions at all momenta scale
while the latter is a zero momentum observable. In our specific
ansatz for the hidden sector, the SUSY breaking fields are really
decoupled from the current multiplets of $G_{SM}^{(1)}$ for $p\gg
v$. This effectively means that the integral~\scalarsGGM\ is
cutoff at $p\sim v$. This in turn is responsible for a parametric
suppression of the scalar masses if $v\ll M$.\foot{A corollary
from this description is that the leading order in $v^2/M^2$
depends only on the zero-momentum value of
$C_{0}-4C_{1/2}+3C_{1}$.} On the other hand, the gaugino mass is
given by a zero momentum correlation function, and thus it is
oblivious to all the complicated symmetry breaking patterns at
nonzero energy.

Physically we can think of these models as giving the messenger
fields charge under the MSSM only at energy scales below~$v$. In
other words, we effectively assign charges which are energy
dependent. This does not affect the gauginos but creates a
suppression for the scalars (under the conditions we discussed).
In section~4 we mention some of the many possible phenomenological
consequences.

It is important to keep in mind that the MSSM RG flow gives rise
to a contribution to the soft scalar masses of the following form
\eqn\gauginocont{\delta m_{\tilde f}^2\sim {\alpha\over
4\pi}{m_{\tilde g}^2}~.} This can be interpreted as a contribution
to the soft scalar masses at three loops and it is further
logarithmically enhanced if the mediation scale is high. Hence,
unless some tuning is invoked, we do not expect to be able to make
the gauginos of the MSSM heavier than the scalars by more than a
factor of a few. The details depend on the species of particles we
are interested in. Thus, the practically interesting values of
$v/M$ cannot be much smaller than $\CO(10^{-1})$.

While the ansatz of Fig.1 is interesting, it seems baroque. The
suppression of the scalar masses is highly dependent on the fact
that the link fields are supersymmetric for vanishing
$G_{SM}^{(2)}$ gauge coupling. Further, no fields in the
SUSY-breaking sector can be charged under $G_{SM}^{(1)}$. To get
the suppression we also need to require the higgsing scale $v$ to
be smaller than the messenger scale $M$. Since eventually we need
to look for dynamical models, preferably as simple as possible,
the natural question is whether the involved structure of Fig.1
can emerge from the dynamics of some ``single sector'' model. This
may seem unlikely because in such models all the particles
interact strongly with all the others. Then, we expect that all
the fields feel SUSY breaking equally.

This intuition is indeed correct in many cases, and the scalar
suppression is washed out. But there is a large class of
single-sector theories which can be analyzed and shown to lead to
spectra with $m_{\tilde f}\ll m_{\tilde g}$. We give such examples
in the next section and explain why the mechanism we present is
generic.

%%%%%%%%%%%%%%%%%%%%%%%%%%%%%%%%%%%%%%%%%%%%%%%%%%%%%%%%%%%%%%%%%%%
%%%%%%%%%%%%%%%%%%%%%%%%%%%%%%%%%%%%%%%%%%%%%%%%%%%%%%%%%%%%%%%%%%%

\newsec{Dynamical Models}

The main conceptual ingredient we will utilize is that higgsing in
the magnetic low energy effective theory corresponds to adding a
mass term in the electric theory. As we saw in the previous
section, higgsing is an important ingredient in realizing this
sfermion mass screening mechanism. In addition, adding mass terms
in the UV gives rise generically to metastable SUSY breaking
vacua. This is why, in fact, many dynamical models can lead in the
IR to both SUSY breaking and suppressed scalar masses.

Of course, we have to look for models with an appropriate sector
of link fields. As we explained, our increased understanding of
these strongly coupled theories has shown that such features, even
though they may not seem intuitive, can arise generically. This
does not mean that they always occur. Indeed, in the simplest
version of massive SQCD, it seems impossible to find a mechanism
to suppress the scalar masses.\foot{The main trouble is the fact
that we typically get messengers charged under the SUSY-breaking
flavor group directly. In other words, there is no good sub-sector
of link fields.} However, as we will see next, massive SQCD with
vector like Yukawa couplings has an effective theory which drives
spontaneous SUSY breaking, on top of the structure we presented in
section~2.

%%%%%%%%%%%%%%%%%%%%%%%%%%%%%%%%%%%%%%%%%%%%%%%%%%%%%%%%%%%%%%%%

\subsec{Massive SQCD with Yukawa Interactions} Consider $SU(N_c)$
SQCD with $N_f$ quarks $Q^I$ in the fundamental representation and
$N_f$ quarks $\tilde Q_I$ in the anti-fundamental representation.
We divide them into two subsets of $k$ and $N_f-k$ quarks, labeled
with the indices $i=1...k$ and $a=k+1...N_f$. We also introduce
some singlets $S_i^a$ and $\tilde S^i_a$. The theory is simply
given by\foot{Some aspects of SQCD with singlets coupled to the
quarks were analyzed in~\BarnesJJ. Possible applications of this
theory were discussed, for example,
in~\refs{\FrancoWF\CraigHF\AmaritiSZ-\SchaferNamekiIZ}. However,
in our case the singlets are not strictly necessary; the Yukawa
couplings can be replaced by dangerously irrelevant operators
composed out of electric quarks.} \eqn\lag{W_{electric}=m_{I}^J
Q^I\tilde Q_J+S_i^a Q^i\tilde Q_a+\tilde S^i_a Q^a\tilde Q_i~.} To
preserve a large flavor symmetry group we choose the mass matrix
to be $m_{I}^J=m_1\delta_{i}^j\oplus m_2\delta _a^b$, so quarks in
the first set have mass $m_1$ and in the second set $m_2$. If we
also include a mass term $m_S S^i_a \tilde S^a_i$, then this
theory is the most general renormalizable theory respecting the
$SU(k)\times SU(N_f-k)$ flavor group. For now we treat this group
as global (although later we will gauge parts of it).

The theory also has two abelian symmetries. One is the usual
baryon number $U(1)_B$. The second assigns charge $\pm 1$ to
$Q^{i},\tilde Q_i$ respectively and $\pm1$ to $\tilde S^i_a,S_i^a
$. We denote this symmetry by $U(1)'$. These symmetries are not
anomalous. They also have no cubic anomalies.

When all the mass terms are set to zero the theory has a new
global symmetry. It is therefore natural to imagine these masses
are small (compared to the strong coupling scale). In this regime
the mass terms for the quarks are physically crucial, but one can
set $m_S=0$ without changing any of the essential details of the
analysis. This is why we do not include the mass term explicitly
in~\lag\ and do not discuss it any further.

For $N_f<3N_c$ the gauge coupling grows strong in the IR and all
the matter fields interact strongly with each other. In spite of
that, we will show (for some regime of parameters) that this
theory in fact reduces precisely to the ansatz of the previous
section. We will find that the scales $m_1$, $m_2$ are related to
$v$, $M$.

We concentrate on $N_f<3N_c/2$, where at low energies and small
VEVs, the theory flows to the IR-free Seiberg dual
description.\foot{We are careless about the distinction between
the electric and magnetic strong scales. For simplicity, we assume
that they are identical and drop all the incalculable numbers in
the dual description.} The superpotential is
\eqn\seibergdual{W_{magnetic}=qM\tilde q+\Lambda
S^a_iM^i_a+\Lambda\tilde S^i_a M_i^a+\Lambda Tr(mM)~.} The
magnetic gauge group is $SU(N_f-N_c)$. The magnetic quarks
$q,\tilde q$ are charged, while the mesons $M$ and the original
singlets $S,\tilde S$ are neutral.

We take $m_{1,2} \ll \Lambda$ and denote
$\mu_{1,2}^2=-m_{1,2}\Lambda$. The theory~\seibergdual\ is weakly
coupled.  We see that the mesons $M_i^a$, $M^i_a$ are massive with
mass of order $\Lambda$ and can be integrated out. To write the
resulting action we parametrize the $N_f$ magnetic quarks as
$q=(\chi_i,\psi_a)$, $\tilde q=(\tilde\chi^i,\tilde \psi^a)$.
After integrating out the heavy states, the theory at low energies
is \eqn\finthe{W=\chi N\tilde \chi+\psi M\tilde \psi
-\mu_1^2N_i^i-\mu_2^2M^a_a ~.} In the above we have decomposed the
meson such that the upper block is denoted $N$ and the lower one
$M$ \eqn\mesonde{M_{N_f\times N_f}=\left(\matrix{N_{k\times k} & 0
\cr 0 & M_{(N_f-k)\times (N_f-k)}}\right)~.} The off-diagonal
components were integrated out using the equations of motion of
the $S,\tilde S$ superfields. Therefore, the low energy effective
theory consists of two sectors which interact arbitrarily weakly
at the IR, through the magnetic gauge fields. This is an emergent
phenomenon.

The analysis of both of these sectors is straightforward. To make
the discussion even simpler we choose $k=N_f-N_c\equiv N$,
although analogous results hold as long as either $k\leq N$ or
$k\geq N_c$.

Let us start from analyzing the sector of the $\psi$ quarks. There
are $N_f-N=N_c$ flavors of $\psi$ quarks. Since in the free
magnetic phase $N_c>N$, we cannot satisfy the $F$-term equations
for the $M_a^b$ mesons and SUSY is broken for small field VEVs.
Note the similarity of this sector to the ISS model~\refs{\IntriligatorDD},
except
that the number of quarks is $N_c$ rather than $N_f$. We will see
that although this sector is indeed similar, other aspects of the
model are qualitatively different.

It is convenient to further decompose the quarks and mesons as
follows \eqn\fdecom{M=\left(\matrix{X_{N_f-N_c} & Y \cr \tilde Y &
Z_{2N_c-N_f}}\right)~,\qquad
\psi=(\lambda_{N_f-N_c},\rho_{2N_c-N_f})~,\qquad \tilde
\psi=(\tilde \lambda_{N_f-N_c},\tilde \rho_{2N_c-N_f})~.} In the
attempt to cancel as many $F$-terms as possible we expand around
$X=0, \lambda=\tilde\lambda=\mu_2\unit$. This breaks the
symmetries as follows \eqn\fs{SU(N)_\chi\times SU(N_c)_\psi\times
[SU(N)]\hookrightarrow SU(N)_\chi\times SU(2N_c-N_f)_\rho\times
SU(N)_{diagonal}~.} We have added subscripts to the various
symmetry groups to make it obvious how they act.
$SU(N)_{diagonal}$ is a diagonal combination of the magnetic group
and the $SU(N)_{\lambda}$ flavor group embedded in the original
$SU(N_c)_\psi$. The fact that this $SU(N)_{diagonal}$ leaves the
vacuum invariant and mixes with the magnetic group will have far
reaching consequences. This is the seed for the higgsing
phenomenon we need to realize the scenario outlined in section 2.
Note that the magnetic gauge group is emergent, so this seed is
not even visible in the original description of the theory.

To analyze the spectrum of this vacuum we first turn off the
magnetic gauge group and treat it as a global symmetry. First of
all, we find all the Nambu-Goldstone bosons associated to this
breaking. There are also pseudo-moduli
$Re(\lambda-\tilde\lambda)$ and $Z$. The modes in
$Re(\lambda-\tilde\lambda)$ obtain positive mass squared at
one-loop and are thus set to the origin. Similarly, a one-loop
Coleman Weinberg potential sets $Z=0$ as well. Therefore, this is
a good SUSY-breaking vacuum, with no noncompact massless fields.
Reintroducing the magnetic gauge coupling, we find that some of
the Goldstone bosons are eaten and that some of the pseudo-moduli
obtain an additional positive mass-squared contribution. The
vacuum above is obviously $D$-flat.

We now turn to analyzing the sector of the $\chi$ quarks. The rank
of the matrix $\chi\tilde\chi$ is at most $N_f-N_c$, and this can
be arranged to cancel all the $F$-terms of $N_i^j$,
\eqn\chisector{\langle\chi\tilde \chi\rangle=\unit_{N} \mu_1^2~.}
We can use the unbroken $SU(N)_\chi\times SU(N)_{diagonal}$ to
diagonalize $\chi$, $\chi=\mu_1diag(a_i)$, which guarantees
by~\chisector\ that $\tilde \chi$ is diagonal as well
$\tilde\chi=\mu_1diag(a^{-1}_i)$. Further, the energy is minimized
as long as the $D$-terms of the magnetic group vanish. This is
achieved if $a_i=a$ for some general complex $a$.\foot{We are
using the magnetic $D$-terms to conclude that $a_i=a$ for all $i$,
but one could worry that it is possible to involve the $\chi$
quarks too and find more complicated vacua. This does not happen
because of the one-loop positive masses the $\chi$ quarks obtain,
they are fixed to the VEV we discussed after~\fdecom.} This last
remaining noncompact modulus is fixed to $a=1$ if we gauge the
vector like $U(1)'$ symmetry. Thus we summarize that our vacuum is
\eqn\chisectori{\langle\chi\rangle=\mu_1\unit_N~,\qquad
\langle\tilde \chi\rangle= \mu_1\unit_{N}~.} This vacuum further
breaks the symmetry~\fs\ \eqn\symmbreaki{SU(N)_\chi\times
SU(2N_c-N_f)_\rho\times SU(N)_{diagonal}\hookrightarrow
SU(N)_{visible}\times SU(2N_c-N_f)_\rho~.} Note that this last
step contains another crucial ingredient for achieving the
mechanism of section 2; the breaking pattern via the $\chi$ quarks
(which are supersymmetric at tree-level) is of the form
$SU(N)\times SU(N)\hookrightarrow SU(N)$.

Therefore, somewhat surprisingly, the simple theory~\lag\ has a
very rich structure in the IR, accommodating all the necessary
ingredients for suppressed scalar masses. This theory, in spite of
all its components being strongly interacting at some intermediate
scale, reduces in the IR to two weakly interacting sectors, one
breaks SUSY and the other does not. These sectors communicate via
the magnetic gauge fields. The mass terms in the UV trigger
higgsing of the magnetic group. This crucial feature is a general
phenomenon in theories with electric-magnetic duality. The flavor
symmetries mix with the magnetic group generators and eventually
break further to diagonal combinations.

The dictionary between this model and section~2 is
straightforward. We can weakly gauge\foot{Before gauging any
flavor symmetries, there are degenerate vacua where $\lambda=
\tilde \lambda = 0$ and $\tilde \rho_i = \rho_i \neq 0$. These
vacua do not give the desired pattern of higgsing. Even though,
strictly speaking, for our vacuum to exist this degeneracy does
not have to be lifted, it is automatically removed by several
mechanisms which are described in section~4.} $SU(N)_\chi\times
SU(N)_\lambda$ and let the matter fields of the MSSM carry just
$SU(N)_\chi$ charges (of course the simplest choice is to take
$N=5$). The SUSY breaking scale (which is roughly also the
messenger mass in this model) is $\mu_2$ while the higgsing scale
is $\mu_1$. The link fields are the $\chi$ magnetic quarks. Hence,
the scalar mass squared will be suppressed roughly by
$\mu_1^2/\mu_2^2$ relative to the naive value. For convenience, we
summarize the dictionary between the deconstruction ansatz and the
dynamical model in the following table: \thicksize=1pt \vskip12pt
\begintable
Deconstruction | Dynamical Model \crthick gauge group
$G_{SM}^{(1)}$ | $SU(N)_{\chi}$ \cr gauge group $G_{SM}^{(2)}$ |
$SU(N)_{\lambda}$ \cr SUSY breaking scale $M$ |
$\sqrt{m_2\Lambda}$ \cr higgsing scale $v$ | $\sqrt{m_1\Lambda}$
\cr link fields | $\chi, \ \tilde \chi$
\endtable
\noindent Recall that, in the magnetic duality frame, the global
symmetry $SU(N)_{\lambda}$ mixes with the magnetic gauge bosons to
become $SU(N)_{diagonal}$, which in the end of the day plays the
role of $G_{SM}^{(2)}$.

The ratio of $\mu_1^2 / \mu_2^2$ cannot be made arbitrarily small.
The main obstruction (common to all deconstruction-like theories)
is that the $\chi$ quarks receive some soft masses through the
magnetic gauge fields. These should not compete with the
dynamics leading to spontaneous symmetry breaking, so $\mu_1^2 /
\mu_2^2 > 10^{-4}$. As we have explained in section~2, this
constraint is irrelevant from a practical point of view and can
still accommodate any conceivable scalar suppression.

The SUSY-breaking vacuum we found here is metastable.
Supersymmetric vacua are restored via non-perturbative effects
which we discuss in the next subsection.

The construction we presented here is strongly dependent on the
mixing of flavor symmetries with the magnetic gauge group. (This
should perhaps be called ``magnetic color flavor  locking.'')
Since this is a general property of massive theories with a dual
description, our model is in fact easily generalizable to many
other examples. In subsection~3.3 we briefly discuss some
analogous $SO(N)/Sp(N)$ theories.

As it stands, the theory at hand has an accidental unbroken
$R$-symmetry in the IR and so this scalar suppression is not of
much use phenomenologically. This is a minor technical issue and
there are many ways around it. We list some of the simplest ones
in section 4, where we also make some rough estimates of the soft
scalar and gaugino masses in this theory.

%%%%%%%%%%%%%%%%%%%%%%%%%%%%%%%%%%%%%%%%%%%%%%%%%%%%%%%%%%%%%%%%

\subsec{Non-Perturbative Restoration of Supersymmetry}

The search for SUSY vacua can be carried either in the electric or
in the magnetic description, with identical results. Here we will
only discuss the electric description. Furthermore, since SUSY
vacua are protected by holomorphy we are allowed to study~\lag\
with masses $m\gg \Lambda$ and then analytically continue the
results to the regime of interest. (We again set the mass term for
the singlets to zero for simplicity.) This theory is conceptually
easy to understand. Below the scale $\sim m$ it flows to pure SYM
theory and singlets. The scale of this pure SYM theory depends on
the singlets as
\eqn\scama{\Lambda_{eff}^{3N_c}=\det\CM\Lambda^{3N_c-N_f}~,} where
\eqn\massmatrix{\CM=\left(\matrix{m_1\unit_{k} & S \cr \tilde  S &
m_2\unit_{N_f-k}}\right)~.} This creates a superpotential due to
gaugino condensation
\eqn\lowStheory{W_{eff}\sim\left(\det\CM\right)^{{1\over
N_c}}\Lambda^{3-{N_f\over N_c}}~.} The VEVs of the mesons depend
on $S$ and are given by \eqn\mesvevs{\langle Q\tilde Q\rangle\sim
\left(\det\CM\Lambda^{3N_c-N_f}\right)^{{1\over N_c}}\CM^{-1}~.}

As mentioned before, some of the interesting values of $k$ are
$k\leq N_f-N_c$. One simple property of the determinant
of~\massmatrix\ is that this is a polynomial in $S^a_i\tilde
S_b^j$ of order $k$. (The fact that $S$ always appears together
with $\tilde S$ follows from symmetry.) SUSY vacua at finite
distance in field space appear whenever
$\del_S\det\CM=\del_{\tilde S}\det\CM=0$. This is automatically
satisfied for $S=\tilde S=0$ and at some other isolated points at
which $S,\tilde S\sim m$. In all of these cases the meson VEV is
of order $\langle Q\tilde Q\rangle\sim m^{{N_f\over
N_c}-1}\Lambda^{3-{N_f\over N_c}}$ and once written in terms of
the magnetic variables it is clearly parametrically far away from
the origin for $\mu\ll\Lambda$. Lastly, because $k\leq N_f-N_c$
there is a runaway behavior for large $S,\tilde S$ (for example in
the direction where they are all equal). Indeed, for large $S$ the
superpotential~\lowStheory\ takes the form
$W_{eff}\sim\left(m^{N_f-2k}S^{2k}\right)^{{1\over
N_c}}\Lambda^{3-{N_f\over N_c}}$.  (With a mass term for the
singlets this runaway is stabilized.) On this branch some of the
mesons are large and some are small.

Generally speaking, the fact that all the SUSY vacua and the
runaway are far in field space is not surprising. This follows
merely from the fact that these effects can be described by
non-perturbative physics in the IR-free dual gauge theory. They
must, therefore, not affect the structure of the theory at small
VEVs.

%%%%%%%%%%%%%%%%%%%%%%%%%%%%%%%%%%%%%%%%%%%%%%%%%%%%%%%%%%%%%%%

\subsec{Orthogonal and Symplectic Groups} First, we will consider $SO(N_c)$
gauge theory with $N_f$ flavors, $Q_I$ with $I=1..N_f$, in free
magnetic range, namely $N_c-2 < N_f < {3\over 2}(N_c - 2)$. As in
the $SU(N)$ case, we divide these electric quarks into two subsets,
including $k$ and $N_f-k$ quarks labeled by the indices $i$ and
$a$. Now we add singlets under the gauge group $S_{ia}$ and
consider the following superpotential (with $m_{IJ}$ being a
full-rank symmetric matrix) \eqn\SOelec{W  = m_{IJ} \vec Q_I \cdot
\vec Q_J + \lambda S_{ia} \vec Q_i \cdot \vec Q_a~.} We have not
included a mass term for $S$ although this is needed to make the
theory generic. The reason is that, like in the case of $SU(N_c)$,
including it does not affect the results of the analysis. The same
comment holds for our $Sp$ theory.

In the IR, the theory is described by its Seiberg dual with a magnetic
group $SO(N \equiv N_f-N_c +4)$~\IntriligatorID. The global
symmetry of the model is $SO(k) \times SO(N_f -k)$. If the
coupling $\lambda$ in~\SOelec\ vanishes and the matrix $m_{IJ}$ is
proportional to the identity, the symmetry is enhanced to
$SO(N_f)$. To further simplify the discussion we again take $k=N$
and in order to preserve a global symmetry $SO(N) \times SO(N_c -
4)$ we choose the quark mass matrix to be $m_{IJ} = m_1
\delta_{ij} \oplus\ m_2 \delta_{ab}$. In the IR, the
superpotential is \eqn\SOmagn{W = q^T M q - \mu_1^2 M_{ii} -
\mu_2^2 M_{aa} + \lambda \Lambda S_{ai} M_{ai}~.} Here the
magnetic meson $M$ is a symmetric $N_f \times N_f$ matrix and
unlike in the $SU(N)$ case there are, of course, no conjugate
quarks.

Assuming that $\lambda \sim 1$ and integrating out the heavy
fields $S_{ai}$ and $M_{ia}$ we find an IR effective theory which
is very similar to what we had in the case of $SU(N)$. The
SUSY-breaking vacuum of the model is obtained for \eqn\SOvac{ q^T
= \left( \matrix{\mu_1 \unit_{N\times N}, \mu_{2} \unit_{N\times
N}, 0} \right), \ \ \ M_{IJ} = 0~.}
The vacuum
energy of this state is $(2N_c - N_f - 8)\mu_2^4$. At the scale
$\mu_2$, the symmetry is broken as follows (as before, we
decompose the magnetic quarks as $q=(\chi,\lambda)$):
\eqn\sbt{SO(N)_{\chi} \times [SO(N_c-4)\times SO(N)_{mag}]
\hookrightarrow SO(N)_{\chi} \times SO(N)_{diagonal} \times
SO(2N_c- N_f - 8)} At the scale $\mu_1$ this global symmetry is
further broken down to \eqn\sbo{SO(N)_{\chi} \times
SO(N)_{diagonal} \times SO(2N_c- N_f - 8) \hookrightarrow
SO(N)_{visible} \times SO(2N_c - N_f -8)} Analogously to the
$SU(N)$ theory, there are some Nambu-Goldstone bosons and some
pseudo-moduli (which are stabilized at one-loop).

The situation is also very similar in the $Sp(N_c)$ theory with
$2N_f$ electric quarks in the fundamental representation. The free
magnetic phase occurs for $(N_c + 3) \le N_f < {3\over 2} (N_c
+1)$.\foot{In this analysis we follow the conventions
of~\IntriligatorDD.} We again divide our quarks into two sets,
which include $2N$ and $2(N_f-N)$ quarks, where $N \equiv N_f-N_c
-2 $. Introducing new gauge singlets $S_{ai}$ we postulate the
following superpotential \eqn\SPW{W = -m_{IJ}Q^I_{\alpha}
Q^J_{\beta}J_{\alpha \beta} + \lambda S_{ia}Q^i_{\alpha}
Q^a_{\beta} J_{\alpha \beta}~. } The breaking occurs largely along
the lines of $SO(N)$ and $SU(N)$ and will not repeat all the steps
here. However, we note that the global symmetry of the model with
the superpotential~\SPW\ is $Sp(N) \times Sp(N_f-N)$. In the
minimum of the IR effective theory this group is spontaneously
broken down to $Sp(N)_{visible} \times Sp(N_f-2N)$. This is
triggered again by mixing with the magnetic gauge group, as in the
previous two cases.

%%%%%%%%%%%%%%%%%%%%%%%%%%%%%%%%%%%%%%%%%%%%%%%%%%%%%%%%%%%%%%%%%%%
%%%%%%%%%%%%%%%%%%%%%%%%%%%%%%%%%%%%%%%%%%%%%%%%%%%%%%%%%%%%%%%%%%

\newsec{Comments on Phenomenology}

This section deals with two closely related topics. We first take
a closer look at the phenomenology of the dynamical
models we constructed above. Specifically, we will discuss various possibilities for
breaking the accidental $R$-symmetry and roughly estimate some
soft masses.

In the second part of this section we attempt to obtain a broader
view of the phenomenological possibilities. Recall that the
investigation of models of messengers teaches us a lot about gauge
mediation, and in the same spirit, the analysis of models such as
those in section~2 should be a good guide to many dynamical models
of gaugino mediation (including those we discuss in this paper).
We will see that many novel phenomenological scenarios are
naturally embedded in such theories.

%%%%%%%%%%%%%%%%%%%%%%%%%%%%%%%%%%%%%%%%%%%%%%%%%%%%%%%%%%%%%%%%%%%

\subsec{Accidental $R$-Symmetry Breaking and Soft Masses} Consider
the $SU(N)$ theory we described in subsection~3.1. For simplicity
we choose $N=5$. We identify $SU(5)_{\chi}$ with the GUT group.
Then we notice that $SU(5)_{diagonal}$ from~\fs\ plays the role of
$G_2^{SM}$ from section~2, and the fields $\chi, \tilde \chi$ are
the link fields. The model we discuss has an accidental
$R$-symmetry which is not broken in the meta-stable vacuum, and
therefore, gaugino masses do not emerge. To render the model
realistic this accidental symmetry should be broken (spontaneously
or explicitly). Breaking this symmetry is a technical issue and
there are many ways to do it (relevant references can be found in
the review~\KitanoFA). For instance, small quartic deformations in
the UV become relevant in the IR and can easily lead to
$R$-symmetry breaking, either spontaneous or
explicit.\foot{Various models of messengers with $R$-symmetry
breaking effects have been studied in some generality
in~\MarquesVA.}

To be concrete let us focus on adding a quartic deformation in the
UV, which in the IR becomes
\eqn\quartdef{\delta W = {\epsilon
\mu_2 \over 2} \left(\Tr Z^2 \right)~.}
This deformation was
studied in detail, for example, in~\refs{\GiveonEF,\EssigKZ}. One
finds that~\quartdef\ shifts the VEV of $Z$ away from the origin
to\foot{One could also imagine a double-trace deformation $(\Tr
Z)^2$, but this does not affect the discussion qualitatively, for
a dedicated study the reader is referred to~\EssigKZ.}
\eqn\RutgZ{Z
\sim {\epsilon \mu_2\over (\alpha_h/4\pi)} \unit~.}
In this
formula $\alpha_h=h^2/4\pi$ with $h$ being the Yukawa coupling in
the Seiberg dual. We have set this coupling to one throughout the
paper but in this formula we reintroduce it in order to make the
parametric dependence manifest. In practice, the constraints of
metastability and the structure of the one loop effective
potential do not allow us to probe arbitrary values of $Z$.

The main point here is that in spite of the fact that the
deformation~\quartdef\ breaks the $R$-symmetry explicitly, the
spontaneous breaking through $\langle Z \rangle$ is larger by an
inverse loop factor. Thus, effectively, we get spontaneous
breaking. This point is of conceptual importance, because by the
Nelson-Seiberg theorem~\NelsonNF, explicit breaking would be in
tension with metastability~\refs{\IntriligatorCP,\IntriligatorPY}.
Indeed, in realizations such as~\RutgZ, new vacua emerge in the IR
effective theory, but the transition rate between our vacuum and
these new vacua is parametrically suppressed.

There are several other ideas which lead to very similar effective
descriptions. For example, if massless electric quarks are
introduced, we can use the mechanism of~\GiveonNE\ to probe large
values of $\langle Z\rangle$ as well. To list a few other options,
there is the baryonic deformation of~\AbelJX\ or models like
in~\refs{\DineXT,\HabaRJ}.

Because the breaking is mostly spontaneous, we can understand the
physics without going example by example.  We study the theory as a
function of the pseudo-modulus VEV
$\langle Z\rangle $. The phenomenological predictions then depend
on $\langle Z\rangle$ and can be translated immediately to
microscopic realizations upon expressing the VEV in terms of the
microscopic variables.

It is well know that gaugino masses as a function of $\langle
Z\rangle $ are only formed at the subleading order in
SUSY-breaking parameter. Our simple model does not have a strict
small SUSY breaking expansion parameter. (It can be easily
introduced by further allowing the electric quarks have different
masses, but we will not carry out this analysis here.) The absence
of a leading order contribution to the gaugino mass is curiously
reflected in a numerical $\sim 0.1$ suppression. The rough
estimate of the soft gaugino mass as a function of $\langle
Z\rangle$ is then \eqn\gaugmas{m_{\tilde g}\sim 0.1{\alpha\over
4\pi}\Biggl\{ \matrix{\mu_2^6/ \langle Z\rangle^5 &~, \ \ \langle
Z\rangle \gg\mu_2\cr \mu_2 & ~,\  \ \langle Z\rangle \sim  \mu_2
\cr \langle Z\rangle & ~,\  \ \langle Z\rangle \ll \mu_2}} This
rather small contribution to the gaugino soft mass posed severe
phenomenological troubles in gauge-mediation model-building based
on ISS. However, here it is less of a problem because the scalar
masses can be screened as well. There are also models which
produce gaugino masses at leading order, as in~\softmasses,
including~\refs{\KitanoXG,\GiveonYU}. We will not elaborate on
these models, but we would like to note that we believe many of
their phenomenological features can be drastically improved in our
framework.

We now switch to discussing the soft scalar masses. The two-loop
contribution has the $v^2/M^2$ suppression factor we discussed at
length in section~2, and in our model the result can be estimated
as \eqn\scamas{m^2_{\tilde f}\sim \left(\alpha\over 4\pi\right)^2
\mu_1^2~.} This estimate of the two-loop contribution holds when
$\mu_1^2 \ll \mu_2^2$.  It is independent of $\langle Z\rangle $
and of the ratio $\mu_1 \over \mu_2$ due to a peculiar see-saw
like messenger structure for $\langle Z\rangle \gg
\mu_2$.\foot{This peculiarity, in fact, appears in many O'R-like
models. It was systematically studied in~\KatzGH.} In more detail,
since we have chosen to study the simplest version of the model,
we find that, for all $\langle Z\rangle$, the messenger masses and
mass splittings are comparable. Equation~\scamas\ can therefore be
understood as a result of the usual screening formula
 $m^2_{\tilde f} \sim \left(\alpha\over 4\pi\right)^2{\mu_1^2 \over M^2}{F^2 \over M^2}$
 with $F \sim M^2$.  Of course, since $\mu_1$ is smaller than all the other parameters in the
problem,~\scamas\ reflects a screened contribution to the scalar
masses.

Very importantly, there are also three-loop contributions which
could be even larger than~\scamas, depending on the parameters.
The idea is that the $\chi$ quarks have some soft masses at
two-loops, roughly given by ${\alpha^2\over 16\pi^2}\mu_2^2$.
These $\chi$ quarks then play the role of the scalar component of
a massive gauge multiplet, and therefore this feeds to the visible
particles. This contribution is estimated as
\eqn\threeloops{m^2_{\tilde f}\sim{\alpha^3\over (4\pi)^3}
\mu_2^2~.} This estimate is applicable as long as
$\mu_1^2>{\alpha^2\over 16\pi^2}\mu_2^2$, which is definitely the
regime one should be interested in anyway. A more careful analysis
also shows that there is no logarithmic enhancement
of~\threeloops\ but we will not go into details here since this
will take us too far afield (see~\DeSimoneGM\ for a related
discussion). We see that depending on the comparison of
$\mu_1^2/\mu_2^2$ to a loop factor, this can be negligible or
dominant compared to~\scamas. More interestingly, if the gaugino
mass does not arise at leading order, we expect~\threeloops\ to be
more important than the usual gaugino mediated
contribution~\gauginocont.

Another parenthetical issue about this specific model is that some
of the $\lambda, \tilde \lambda $ scalars are massless. Since
those particles act as messengers and are charged under the SM this
feature should be removed. This particular problem is trivially
fixed, often by the same source that breaks $R$-symmetry. In
addition, these massless scalars automatically disappear once we
gauge flavor symmetries. Lastly, these massless particles
disappear if we allow a more general mass matrix in the UV
Lagrangian. (This is also a useful way to construct theories which
are not necessarily at some relatively low scale.)

Finally, we would like to note that dynamical models with large
flavor symmetries and which are naturally perturbative up to the
GUT scale are not known, and our model does not change this state
of affairs. However, theories which reduce to the deconstructed
ansatz are expected to have a smaller beta function than usual.
Indeed, models based on ISS like those we tried to encompass in
the analysis above, when fitted into our framework, generally
behave better than their original incarnations.~\foot{In
deconstructed models the gauge couplings are shifted by the Higgs
mechanism. Unless this is done in an $SU(5)$ invariant way (or
some other special relations are satisfied) perturbative
unification is ruined. This is similar to constraints on the
messenger fields in ordinary models of gauge mediation.}

%%%%%%%%%%%%%%%%%%%%%%%%%%%%%%%%%%%%%%%%%%%%%%%%%%%%%%%%%%%%%%%%%%%%%%%%%%%%%%%%%%%%%%%%%%%%%

\subsec{Spectroscopy}

Even the simple theory we discussed above, as a function of its
parameters and the choice of how to break the $R$-symmetry, can
give rise to a diverse spectrum. It varies from a theory with an
unnatural spectrum (the sparticles being significantly heavier
than the gauginos) all the way to some form of gaugino mediation.
In our particular scenario we chose to focus on a theory which is
naturally mediating SUSY breaking at a low scale, but as we
remarked, the mediation scale can trivially be promoted to a free
parameter.
Therefore, we should consider both high and low scale mediation
scenarios.

In order to get some handle over these various phenomenological
scenarios, it is convenient to step back and reconsider the ansatz
of section~2. In particular, we can define its simplest version as
``minimal gaugino mediation.'' It consists of a sector of link
fields which are forced to get VEVs due to a Lagrange multiplier
$\CA$ in the superpotential $\int d^2\theta \CA\left(L\tilde
L-v^2\right)$. The Hidden sector $H$ consists of a SUSY-breaking
spurion $X=M+\theta ^2 F$ which couples to a single pair of
messengers $X\psi\tilde\psi$.

We see that this minimal model has three dimensionful scales, one
more than minimal gauge mediation. As a function of $v/M$ for
large $v$ (say a little below the GUT scale) this model
interpolates between high scale minimal gauge mediation (see the
review~\GiudiceBP) and high scale gaugino mediation (considered,
for example, in~\KaplanAV). For low $v$ (e.g.~$\sim$10~TeV) the
model interpolates between low scale minimal gauge mediation and
low scale gaugino mediation. The phenomenology of the latter
scenario is much less studied, but the analysis
of~\refs{\DeSimoneGM,\DeSimoneWS} shows convincingly that such
theories, even in their minimal form, lead to spectacular collider
signatures.

To the best of our knowledge, the particle spectrum has not been
systematically computed as a function of $v/M$. Given that it
interpolates between such canonical (and compelling) scenarios, it
cannot be overemphasized how important it is to understand the
intermediate regime.

We see that even for ``minimal gaugino mediation'' the
phenomenology (and particle spectrum as a function of the
parameters) is largely unknown, it is then not surprising that the
simplest and most interesting generalizations of it have not been
studied at all. Even though these models fall into the class of
GGM, a lot of the ``common lore'' about gauge mediation can be
violated. As a first example, some of the general results
in~\DumitrescuHA\ about theories of messengers do not hold in our
theories; the bino can be pretty heavy. Further, we can take into
account doublet-triplet splitting in the messenger sector or in
the link field spectra.\foot{Analogous studies in gauge-mediation
scenarios led to novel and surprising collider signatures, see
e.g.~\refs{\CheungES\MeadeQV\KatzQX-\KatzXG}.} We expect that such
models give rise to exotic spectra whose signatures have not been
studied. For instance, it is conceivable that the usual hierarchy
between the sparticles $m_{\tilde q}
> m_{\tilde L} > m_{\tilde e_R}$ can be altered. This phenomenon,
which is not standard regardless of the masses of the gauginos,
can lead to surprising consequences if the gauginos are relatively
heavy. For example, we might find a sneutrino NLSP. Even if the
scalars are not all lighter than the gauginos, a reversed ordering
of the scalars in the spectrum might be very interesting as well.

Models like those of section~2 may have looked unnatural before,
but we believe that the fact that they arise from perfectly
natural (and generic) dynamical mechanisms, makes them very well
motivated phenomenologically and warrants a careful investigation
of their possible phenomenological signatures.

\newsec{Summary and Open Questions}

In this note, we showed that gaugino mediation can arise
dynamically in asymptotically free gauge theories.  Using Seiberg
duality, we were able to find calculable examples where the
screening of soft masses arises through higgsing of the magnetic
gauge group.  We expect calculable examples to be abundant with
low energy effective descriptions given by various gauged
WZ-models resembling Fig.1.

The scalar screening necessary for gaugino mediation has commonly
been considered an extra dimensional or strongly coupled
phenomena. While these examples do arise in the strongly coupled
regime of the ``electric" description, they are fully calculable.
Because these models dynamically give rise to a
deconstruction-like Lagrangian, it suggests that other
``extra-dimensional phenomena'' may also emerge from simple single
sector four dimensional theories. For example, it would be natural
to study models of flavor by modifying the models presented here.

From a more phenomenological perspective, the spectra of particles
in these models obey the rules of general gauge mediation.
However, given a generic model, the resulting spectrum would
appear to be quite unconventional.  Given how naturally these models arise within SUSY gauge theories,
these models warrant further phenomenological study.

We close by listing few of the many open questions which we
believe are worth pursuing. As far as dynamical models are
concerned, we have only discussed the tip of the iceberg of
possibilities.  It should be better understood which theories are
capable of screening the scalar masses and what
common features they share. Our phenomenological knowledge of such theories
is very limited unfortunately; even the spectrum of soft masses in
the minimal effective model is not know as a function of its
basic parameters. There are excellent
reasons to believe that these theories lead to interesting spectra
and novel collider signatures. Therefore, it would be nice to
systematically study such models. As a starting point for such a
systematic study, it would be useful to understand better all the
different kinds of contributions to the visible soft masses.
Finally, it is also tempting to try and embed a mechanism for the
$\mu/B_\mu$ terms in these theories.

\bigskip
\bigskip
\bigskip

\bigskip
\bigskip
\bigskip
\bigskip
\bigskip
\bigskip
\bigskip
\bigskip
\bigskip
\bigskip

\noindent {\bf Acknowledgments:}

We would like to thank R.~Argurio, N.~Craig, T.~Dumitrescu,
A.~Giveon, N.~Seiberg, D.~Shih, and M.~Sudano for comments. The
work of DG was supported in part by DOE grant DE-FG02-90ER40542.
AK was partially supported by NSF grant PHY-0801323. The work of
ZK was supported in part by NSF grant PHY-0503584. Any opinions,
findings, and conclusions or recommendations expressed in this
material are those of the authors and do not necessarily reflect
the views of the National Science Foundation.

\listrefs

\end

%% file: tables.tex
%%%%%%%%%%%%%%%
% +--------------------------------------------------------------------+
% |                                                                    |
% |                           TABLES.TEX                               |
% |                                                                    |
% |                     Ray F. Cowan  15-Feb-85                        |
% |                                                                    |
% |                       Princeton University                         |
% |                                                                    |
% |                     Last Revision: 17-Apr-86                       |
% |                                                                    |
% |   Macros I find handy for making tables.  See TABLEDOC TEX for     |
% |   a longer description.  The token-counting macros are straight    |
% |   from the TeXbook's "Dirty Tricks" appendix.                      |
% |                                                                    |
% +--------------------------------------------------------------------+
%
\newbox\hdbox%
\newcount\hdrows%
\newcount\multispancount%
\newcount\ncase%
\newcount\ncols% This is the number of primary text columns in the table.
\newcount\nrows%
\newcount\nspan%
\newcount\ntemp%
\newdimen\hdsize%
\newdimen\newhdsize%
\newdimen\parasize%
\newdimen\spreadwidth%
\newdimen\thicksize%
\newdimen\thinsize%
\newdimen\tablewidth%
\newif\ifcentertables%
\newif\ifendsize%
\newif\iffirstrow%
\newif\iftableinfo%
\newtoks\dbt%
\newtoks\hdtks%
\newtoks\savetks%
\newtoks\tableLETtokens%
\newtoks\tabletokens%
\newtoks\widthspec%
%
%  Book-keeping stuff--see how often these macros are called.
%
\immediate\write15{%
CP SMSG GJMSINK TEXTABLE --> TABLE MACROS V. 851121 JOB = \jobname%
}%
%
%  Turn on table diagnostics.
%
\tableinfotrue%
\catcode`\@=11%  Allows use of "@" in macro names, like PLAIN.TEX does.
%  Debugging aid.  Writes #1 on the
%                                    user's terminal and in the log file.
%
%  Define the \tstrut height, depth in terms of the x_height parameter.
%
\def\tstrut{\vrule height3.1ex depth1.2ex width0pt}%
\def\and{\char`\&}%  Allows us to get an `&' in the text.  This is the
%                    same as using the PLAIN TeX macro \&.
\def\tablerule{\noalign{\hrule height\thinsize depth0pt}}%
\thicksize=1.5pt%  Default thickness for fat rules.  The user should feel
%                  free to change this to his preference.
\thinsize=0.6pt%   Default thickness for thin rules.
\def\thickrule{\noalign{\hrule height\thicksize depth0pt}}%
\def\ctr#1{\hfil\ #1\hfil}%
%
%
%
%  Here are things for controlling the width of the finished table.
%
\tablewidth=-\maxdimen%
\spreadwidth=-\maxdimen%
\def\tabskipglue{0pt plus 1fil minus 1fil}%
%
%  Stuff for centering or not.
%
\centertablestrue%
%
%
%
%  \vctr vertically centers its argument in the row.
%
\parasize=4in%
\gdef\ARGS{########}%  Produces the correct number of #'s in the preamble
%                      by the time eveything is expanded and \halign sees
%                      it.
\gdef\headerARGS{####}%  Same as \ARGS, but used in \header macros.
\def\@mpersand{&}%  Allows us to get alignment tab characters later
%                   when we have made the character "&" an active macro.
{\catcode`\|=13%  Make |'s locally active.
\gdef\letbarzero{\let|0}%  Globally define a macro that allows us to
%                          keep active |'s from being expanded in edef's.
\gdef\letbartab{\def|{&&}}%
\gdef\letvbbar{\let\vb|}%
%  This \def will cause active |'s read by
%                            \ruledtable to be converted into double
%                            alignment tabs.
}%  End of locally active |'s.
{\catcode`\&=4%  Make these alignment tabs.
\def\ampskip{&\omit\hfil&}%  This local macro skips a vertical rule.
\catcode`\&=13%  Now make &'s into active macros.
\let&0%  This allows us to expand \ampskip in the next \xdef without
%        attempting to expand the & and getting an "undefined control
%        sequence" error.
\xdef\letampskip{\def&{\ampskip}}%
\gdef\letnovbamp{\let\novb&\let\tab&}
%  This will cause active &'s read by
%                                   \ruledtable to be converted into
%                                   double tabs and an \omit'ted \vrule.
}%  End of locally active &'s.
\def\begintable{%  Here we make |'s and &'s active characters so we can
%                  interpret them as macros.  Note that this action is
%                  true only until we encounter the matching \endgroup
%                  token later at the end of the \ruledtable macro.
   \begingroup%
   \catcode`\|=13\letbartab\letvbbar%
   \catcode`\&=13\letampskip\letnovbamp%
   \def\multispan##1{%  We must redefine \multispan to count the number
%                       of primary columns, not physical columns.
      \omit \mscount##1%
      \multiply\mscount\tw@\advance\mscount\m@ne%
      \loop\ifnum\mscount>\@ne \sp@n\repeat%
   }%  End of \multispan macro.
   \def\|{%
      &\omit\widevline&%
   }%
   \ruledtable%  Now we call \ruledtable to do the real work.
}%  End of \begintable macro.
\long\def\ruledtable#1\endtable{%
%
%  This macro reads in the user's data entries
%  and converts them into a ruled table.
%
%  Important note:  Many macros and parameters are re-defined here, and
%  these must be kept local to the table macros to avoid conflict with
%  their use outside of tables.  This is done by the \begingroup token
%  macro \begintable and the \endgroup token at the end of
%  this macro.
%
   \offinterlineskip%  Needed to make rules touch each other.
   \tabskip 0pt%  Needed for same reason as \offinterlineskip.
   \def\widevline{\vrule width\thicksize}%  Make outer \vrule's wider.
   \def\endrow{\@mpersand\omit\hfil\crnorm\@mpersand}%
   \def\crthick{\@mpersand\crnorm\thickrule\@mpersand}%
   \def\crthickneg##1{\@mpersand\crnorm\thickrule
          \noalign{{\skip0=##1\vskip-\skip0}}\@mpersand}%
   \def\crnorule{\@mpersand\crnorm\@mpersand}%
   \def\crnoruleneg##1{\@mpersand\crnorm
          \noalign{{\skip0=##1\vskip-\skip0}}\@mpersand}%
   \let\nr=\crnorule%  A shorter abbreviation.
   \def\endtable{\@mpersand\crnorm\thickrule}%
   \let\crnorm=\cr%  Allows us to use \cr for our own purposes.
%
%  Cause user-typed \cr's to follow a row with a \tablerule.
%
   \edef\cr{\@mpersand\crnorm\tablerule\@mpersand}%
   \def\crneg##1{\@mpersand\crnorm\tablerule
          \noalign{{\skip0=##1\vskip-\skip0}}\@mpersand}%
   \let\ctneg=\crthickneg
   \let\nrneg=\crnoruleneg
   \the\tableLETtokens%  Get the user's extra \let's, if any.
%
%  Put the data entries into a token register so we can scan through them
%  and see what the user is asking us to do.
%
   \tabletokens={&#1}%  We add an extra alignment tab to the beginning
%                       of the first row to allow for the first \vrule.
%
%  Now count how many rows are in the table and return the result in
%  count register \nrows; do the same for columns, and return that
%  in register \ncols.
%
   \countROWS\tabletokens\into\nrows%
   \countCOLS\tabletokens\into\ncols%
%
%  Now do a little arithmetic to convert the number of primary columns
%  into the number of physical columns that the alignment preamble must
%  prepare for;  similarly for rows.
%
   \advance\ncols by -1%
   \divide\ncols by 2%
   \advance\nrows by 1%
%
%  Tell the user how many rows and columns we found in his data, if he
%  wants to know.
%
   \iftableinfo %
      \immediate\write16{[Nrows=\the\nrows, Ncols=\the\ncols]}%
   \fi%
%
%  Now we actually go ahead and produce the table.
%
   \ifcentertables
      \ifhmode \par\fi%  Make sure we are in vertical mode.
      \line{%  The final table comes out as an \hbox of width the \hsize.
      \hss%  The final table will be centered left-to-right.
   \else %
      \hbox{%
   \fi
      \vbox{%
         \makePREAMBLE{\the\ncols}%  Generate the preamble.
         \edef\next{\preamble}%  This line and the next line force the
         \let\preamble=\next%    expansion of all \ARGS tokens into the
%                                appropriate number of #'s.
         \makeTABLE{\preamble}{\tabletokens}%  Go do the \halign here.
      }%  End of \vbox.
      \ifcentertables \hss}\else }\fi%  Finish the centering effect.
%                                       It is important that no spaces
%                                       follow the two `}' here.
%  }%  End of \line.
   \endgroup%  Return all local macros and parameters to their outside
%              values.
   \tablewidth=-\maxdimen%  Reset \tablewidth to normal.
   \spreadwidth=-\maxdimen% Same for \spreadwidth.
}%  End of macro \ruledtable.
\def\makeTABLE#1#2{%  Does an \halign for the \ruledtable macro.
   {%  Start of local parameter values.
   \let\ifmath0%     These macros would cause trouble if they were to be
   \let\header0%     expanded in the following \xdef; we \let them be
   \let\multispan0%  equal to a digit, because digits can't be expanded.
%
%  Set up the width specification here.
%
   \ncase=0%
   \ifdim\tablewidth>-\maxdimen \ncase=1\fi%
   \ifdim\spreadwidth>-\maxdimen \ncase=2\fi%
   \relax%  This \relax is absolutely necessary, without it the following
%           \ifcase will always take \ncase=0.
%
   \ifcase\ncase %
      \widthspec={}%
   \or %
      \widthspec=\expandafter{\expandafter t\expandafter o%
                 \the\tablewidth}%
   \else %
      \widthspec=\expandafter{\expandafter s\expandafter p\expandafter r%
                 \expandafter e\expandafter a\expandafter d%
                 \the\spreadwidth}%
   \fi %
%\out{Widthspec=[\the\widthspec]}%
%\out{Preamble=[\preamble]}%
   \xdef\next{%  We must force the preamble to be expanded BEFORE the
      \halign\the\widthspec{%
%        \halign is done;  this \edef\next{...}\next construction
%                does the trick.
      #1%  This is the preamble text.
      \noalign{\hrule height\thicksize depth0pt}%  Makes the top \hrule.
      \the#2\endtable%  This is the main body.
%
%     \noalign{\hrule height0.7pt depth0pt}%  Makes the last \hrule.
      }%  End of \halign.
   }%  End of \next.
   }%  End of local values.
   \next%  This \next must be outside of the local values, because now
%          we want those troublesome macros in the \let's above to have
%          their normal actions.
}%  End of macro \makeTABLE.
\def\makePREAMBLE#1{%  This macro generates the necessary preamble for a
%                      ruled table with #1 primary columns.
%                      (Primary columns means the number of columns NOT
%                       counting those used for vertical rules.)
   \ncols=#1%  Get the number of columns desired.
   \begingroup%  Start local parameter definitions.
   \let\ARGS=0%  This is the key to the whole thing; it prevents \ARGS
%                from being expanded in the following \edef's.
   \edef\xtp{\widevline\ARGS\tabskip\tabskipglue%
   &\ctr{\ARGS}\tstrut}%  A 1-column preamble.  Gets the sizing right.
   \advance\ncols by -1%  One column has been generated; decrement the
%                         counter.
   \loop%  Append as many further columns as needed to the preamble.
      \ifnum\ncols>0 %
      \advance\ncols by -1%
      \edef\xtp{\xtp&\vrule width\thinsize\ARGS&\ctr{\ARGS}}%
   \repeat
   \xdef\preamble{\xtp&\widevline\ARGS\tabskip0pt%
   \crnorm}%  Adds the last \vrule.
   \endgroup%  End of local parameters.
}%  End of macro \makePREAMBLE.
\def\countROWS#1\into#2{%  This counts the number of rows in #1 by
%                          looking for control sequences that end a row,
%                          e.g., \cr, \crthick, etc., and puts the result
%                          into count register #2.
   \let\countREGISTER=#2%
   \countREGISTER=0%
%  \out{In countROWS:  tokens are [\the#1]}%
   \expandafter\ROWcount\the#1\endcount%
}%
\def\ROWcount{%
   \afterassignment\subROWcount\let\next= %
}%
\def\subROWcount{%
%  \out{In subROWcount:  next is [\meaning\next]}%  Debugging aid.
   \ifx\next\endcount %
      \let\next=\relax%
   \else%
      \ncase=0%
      \ifx\next\cr %
         \global\advance\countREGISTER by 1%
         \ncase=0%
      \fi%
      \ifx\next\endrow %
         \global\advance\countREGISTER by 1%
         \ncase=0%
      \fi%
      \ifx\next\crthick %
         \global\advance\countREGISTER by 1%
         \ncase=0%
      \fi%
      \ifx\next\crnorule %
         \global\advance\countREGISTER by 1%
         \ncase=0%
      \fi%
      \ifx\next\crthickneg %
         \global\advance\countREGISTER by 1%
         \ncase=0%
      \fi%
      \ifx\next\crnoruleneg %
         \global\advance\countREGISTER by 1%
         \ncase=0%
      \fi%
      \ifx\next\crneg %
         \global\advance\countREGISTER by 1%
         \ncase=0%
      \fi%
      \ifx\next\header %
%     \out{In subROWcount:  next=header, ncase set=1}%
         \ncase=1%
      \fi%
%     \out{In subROWcount:  ncase is [\the\ncase]}%
      \relax%
      \ifcase\ncase %
         \let\next\ROWcount%
%        \out{subROWcount---> ncase=\the\ncase}%
      \or %
         \let\next\argROWskip%
%        \out{subROWcount---> ncase=\the\ncase}%
      \else %
      \fi%
   \fi%
%  \out{subROWcount---> NEXT=\meaning\next}%
   \next%
}%  End of macro \subROWcount.
\def\counthdROWS#1\into#2{%
\dvr{10}%
   \let\countREGISTER=#2%
   \countREGISTER=0%
\dvr{11}%
%  \out{In counthdROWS:  tokens are [\the#1]}%
\dvr{13}%
   \expandafter\hdROWcount\the#1\endcount%
\dvr{12}%
}%
\def\hdROWcount{%
   \afterassignment\subhdROWcount\let\next= %
}%
\def\subhdROWcount{%
%\out{In subhdROWcount:  next is [\meaning\next]}%
   \ifx\next\endcount %
      \let\next=\relax%
   \else%
      \ncase=0%
      \ifx\next\cr %
         \global\advance\countREGISTER by 1%
         \ncase=0%
      \fi%
      \ifx\next\endrow %
         \global\advance\countREGISTER by 1%
         \ncase=0%
      \fi%
      \ifx\next\crthick %
         \global\advance\countREGISTER by 1%
         \ncase=0%
      \fi%
      \ifx\next\crnorule %
         \global\advance\countREGISTER by 1%
         \ncase=0%
      \fi%
      \ifx\next\header %
%\out{In subhdROWcount:  next=header, ncase set=1}%
         \ncase=1%
      \fi%
%\out{In subhdROWcount:  ncase is [\the\ncase]}%
\relax%
      \ifcase\ncase %
         \let\next\hdROWcount%
%\out{subhdROWcount---> ncase=\the\ncase}%
      \or%
         \let\next\arghdROWskip%
%\out{subhdROWcount---> ncase=\the\ncase}%
      \else %
      \fi%
   \fi%
%\out{subhdROWcount---> NEXT=\meaning\next}%
   \next%
}%
{\catcode`\|=13\letbartab
\gdef\countCOLS#1\into#2{%
%  \out{In countCOLS:  tokens are [\the#1]}
   \let\countREGISTER=#2%
   \global\countREGISTER=0%
   \global\multispancount=0%
   \global\firstrowtrue
   \expandafter\COLcount\the#1\endcount%
   \global\advance\countREGISTER by 3%
   \global\advance\countREGISTER by -\multispancount
%  \out{countCOLS-->[\the\countREGISTER]}
}%
\gdef\COLcount{%
   \afterassignment\subCOLcount\let\next= %
}%
{\catcode`\&=13%
\gdef\subCOLcount{%
%\out{In subCOLcount: next is [\meaning\next]}
   \ifx\next\endcount %
      \let\next=\relax%
   \else%
      \ncase=0%
      \iffirstrow
         \ifx\next& %
            \global\advance\countREGISTER by 2%
            \ncase=0%
         \fi%
         \ifx\next\span %
            \global\advance\countREGISTER by 1%
            \ncase=0%
         \fi%
         \ifx\next| %
            \global\advance\countREGISTER by 2%
            \ncase=0%
         \fi
         \ifx\next\|
            \global\advance\countREGISTER by 2%
            \ncase=0%
         \fi
         \ifx\next\multispan
            \ncase=1%
            \global\advance\multispancount by 1%
         \fi
         \ifx\next\header
            \ncase=2%
         \fi
         \ifx\next\cr       \global\firstrowfalse \fi
         \ifx\next\endrow   \global\firstrowfalse \fi
         \ifx\next\crthick  \global\firstrowfalse \fi
         \ifx\next\crnorule \global\firstrowfalse \fi
         \ifx\next\crnoruleneg \global\firstrowfalse \fi
         \ifx\next\crthickneg  \global\firstrowfalse \fi
         \ifx\next\crneg       \global\firstrowfalse \fi
      \fi%  End of \iffirstrow.
\relax%\out{subCOL-->  ncase=[\the\ncase]}
% \out{subCOL-->  next=\meaning\next}
      \ifcase\ncase %
         \let\next\COLcount%
      \or %
         \let\next\spancount%
      \or %
         \let\next\argCOLskip%
      \else %
      \fi %
   \fi%
%  \out{subCOL-->  countREGISTER=[\the\countREGISTER]}
   \next%
}%
\gdef\argROWskip#1{%
%  Deletes the next balanced, undelimited argument from a
%                 token list.
% \out{---> Entering argROWskip <---}
% \out{In argROWskip:  deleted arg is [#1]}%
   \let\next\ROWcount \next%
}%  End of macro \argskip.
\gdef\arghdROWskip#1{%
%  Deletes the next balanced, undelimited argument from a
%                 token list.
% \out{---> Entering arghdROWskip <---}
% \out{In arghdROWskip:  deleted arg is [#1]}%
   \let\next\ROWcount \next%
}%  End of macro \arghdROWskip.
\gdef\argCOLskip#1{%
%  Deletes the next balanced, undelimited argument from a
%                 token list.
% \out{---> Entering argCOLskip <---}
% \out{In argCOLskip:  deleted arg is [#1]}%
   \let\next\COLcount \next%
}%  End of macro \argskip.
}%  End of active &'s.
}%  End of active |'s.
\def\spancount#1{%\out{spancount--->\meaning#1}
   \nspan=#1\multiply\nspan by 2\advance\nspan by -1%
   \global\advance \countREGISTER by \nspan
%  \out{number spancount--->\the\nspan; \the\countREGISTER}
   \let\next\COLcount \next}%
\def\dvr#1{\relax}%
% \omit\hfil%
% \parindent=0pt\hsize=1.1in\valign{%
% \vfil#\vfil&\vfil#\vfil\cr\hfil\hbox{\ Added to\ }\hfil&%
% \hfil\hbox{\ empty events\ }\hfil\cr}\hfil%
\def\header#1{%
\dvr{1}{\let\cr=\@mpersand%
\hdtks={#1}%
%\out{In header:  hdtks=[\the\hdtks]}%
\counthdROWS\hdtks\into\hdrows%
\advance\hdrows by 1%
\ifnum\hdrows=0 \hdrows=1 \fi%
%\out{In header:  Nhdrows=[\the\hdrows]}%
\dvr{5}\makehdPREAMBLE{\the\hdrows}%
%\out{In header:  headerpreamble=[\headerpreamble]}%
\dvr{6}\getHDdimen{#1}%
%\out{In header:  hdsize=[\the\hdsize]}%
%\striplastCR{#1}%
{\parindent=0pt\hsize=\hdsize{\let\ifmath0%
\xdef\next{\valign{\headerpreamble #1\crnorm}}}\dvr{7}\next\dvr{8}%
}%
}\dvr{2}}%  End of macro \header.
\def\makehdPREAMBLE#1{%This macro generates the necessary preamble for a
\dvr{3}%
%                      ruled table with \ncols primary columns.
%                      (Primary columns means the number of columns NOT
%                       counting those used for vertical rules.
\hdrows=#1%  Get the number of columns desired.
{%  Start local parameter definitions.
\let\headerARGS=0%
%  This is the key to the whole thing; it prevents \ARGS
\let\cr=\crnorm%
%                from being expanded in the followin \edef's.
\edef\xtp{\vfil\hfil\hbox{\headerARGS}\hfil\vfil}%
\advance\hdrows by -1%  One row has been generated; decrement the
%                         counter.
\loop%  Append as many further rows as needed to the preamble.
\ifnum\hdrows>0%
\advance\hdrows by -1%
\edef\xtp{\xtp&\vfil\hfil\hbox{\headerARGS}\hfil\vfil}%
\repeat%
\xdef\headerpreamble{\xtp\crcr}%
}%  End of local parameters.
\dvr{4}}%  End of \makehdPREAMBLE.
\def\getHDdimen#1{%
%\out{In getHDdimen:  Arg 1=[#1]}%
\hdsize=0pt%
\getsize#1\cr\end\cr%
}%  End of macro getHDdimen.
\def\getsize#1\cr{%
%\out{In getsize:  Arg 1=[#1]}%
%  Here we have to check arg#1 and see if the first token in #1 is an
%    \end; if so, we stop, else we check the width of arg#1.
%  We recall that each arg#1 will be terminated with a \cr token.
\endsizefalse\savetks={#1}%
%\out{In getsize:  the savetks = [\the\savetks]}%
\expandafter\lookend\the\savetks\cr%
%\out{In getsize:  ifendsize = [\meaning\ifendsize]}%
\relax \ifendsize \let\next\relax \else%
\setbox\hdbox=\hbox{#1}\newhdsize=1.0\wd\hdbox%
\ifdim\newhdsize>\hdsize \hdsize=\newhdsize \fi%
%\out{In getsize:  hdsize=[\the\hdsize]}%
%\out{In getsize:  newhdsize=[\the\newhdsize]}%
\let\next\getsize \fi%
\next%
}%
\def\lookend{\afterassignment\sublookend\let\looknext= }%
\def\sublookend{\relax%
%\out{In sublookend:  looknext = [\looknext]}%
\ifx\looknext\cr %
%\out{In sublooknext:  looknext=cr}%
\let\looknext\relax \else %
%\out{In sublooknext:  looknext/=cr}%
   \relax
   \ifx\looknext\end \global\endsizetrue \fi%
   \let\looknext=\lookend%
    \fi \looknext%
}%
%
%  Allow the user to make his own names for crthick, etc.
%
\def\tablelet#1{%
   \tableLETtokens=\expandafter{\the\tableLETtokens #1}%
}%
\catcode`\@=12%  Change @'s back to their normal category code.